\documentclass[a4paper,11pt]{article}
\usepackage{jinstpub} 
\usepackage{lineno}
\usepackage[title,titletoc]{appendix}
\usepackage{comment}
\usepackage{xcolor}
\usepackage{float}
\usepackage{booktabs}
\usepackage{xspace}

\title{3D software compensation of hadronic showers in the CRILIN crystal calorimeter}

\author[a,1]{V.~Ciccarella\note{Corresponding authors}}
\author[a]{E.~Di Meco}
\author[b,1]{R.~Gargiulo}
\author[a]{I.~Sarra}
\author[c]{L.~Sestini}
\affiliation[a]{Laboratori Nazionali di Frascati dell'INFN, Frascati, Italy}
\affiliation[b]{INFN Sezione di Roma1, Rome, Italy}
\affiliation[c]{INFN Sezione di Firenze, Florence, Italy}

\emailAdd{vittoria.ludovica.ciccarella@lnf.infn.it}
\emailAdd{ruben.gargiulo@roma1.infn.it}

\abstract{
Future electron-positron Higgs factories require excellent jet energy resolution to perform precision measurements of Higgs boson couplings to quarks and gluons.
Although homogeneous crystal calorimeters provide remarkable electromagnetic energy resolution, their highly non-compensating response makes hadronic energy reconstruction particularly challenging.
In this work, software compensation techniques are investigated for CRILIN, a longitudinally segmented Cherenkov crystal electromagnetic calorimeter based on PbF$_2$ crystals. Using Geant4 simulations of pion showers, it is shown that shower-shape observables are strongly correlated with the fraction of deposited energy reconstructed in a CRILIN module. Simple event-by-event corrections based on the shower transverse RMS and longitudinal center-of-gravity already yield a substantial improvement in hadronic energy reconstruction. A ParticleNet Graph Neural Network exploiting the full three-dimensional shower topology achieves significantly improved performance with respect to simple energy sum reconstruction. Under several different assumptions for the downstream hadronic calorimeter resolution, the GNN-based reconstruction significantly reduces the effective CRILIN contribution to the combined calorimetric resolution, therefore preserving an excellent combined ECAL+HCAL performance. The dependence of the result on the assumed HCAL resolution is also studied and found to be limited within the range considered. These results demonstrate that highly granular crystal calorimeters can recover a large fraction of the information lost because of their intrinsically non-compensating response through software-based compensation techniques, enabling excellent hadronic energy resolution in a combined ECAL+HCAL system and making such detectors promising candidates for future collider experiments.
}

\begin{document}

\maketitle
\flushbottom

\section{Introduction}

Future electron--positron colliders such as FCC-ee require excellent jet energy resolution to fully exploit their Higgs physics potential and to precisely measure Higgs boson couplings to quarks, with target stochastic terms around $30\%/\sqrt{E[\mathrm{GeV}]}$~\cite{fcc-ee-vol1}. This resolution could be achieved by combining information from the tracking system, the electromagnetic calorimeter (ECAL) and the hadronic calorimeter (HCAL), via the particle-flow technique~\cite{calice}. Nevertheless, to reach this target, a resolution on neutral hadrons of the combined ECAL+HCAL system at the level of $50\%/\sqrt{E}$ or better is needed.

The reconstruction of hadronic showers is fundamentally complicated by fluctuations of the electromagnetic fraction $f_{\mathrm{EM}}$ generated by neutral pion production. In non-compensating calorimeters, where the response to electromagnetic and hadronic energy deposits differs, these fluctuations lead to degraded energy resolution and non-linear response~\cite{wigmans-dual-readout}.

Historically, physical compensation has been achieved through dedicated calorimeter designs combining high-$Z$ absorbers and neutron-sensitive active media~\cite{wigmans-dual-readout}, achieving equal detector responses to the electromagnetic and hadronic energy deposits.
More recently, offline compensation techniques have been developed to estimate $f_{\mathrm{EM}}$ event-by-event and correct the reconstructed energy.
The main offline compensation approach is the so-called dual-readout, which is based on the separate measurement of scintillation and Cherenkov light in an optical calorimeter, which can be mathematically combined to get an estimate of the true deposited energy~\cite{wigmans-dual-readout}. The dual-readout procedure can also be applied to crystal electromagnetic calorimeters upstream of hadronic sections~\cite{lucchini}.
Despite its advantages, it is not technically trivial to build a crystal calorimeter featuring dual-readout, partly because it typically requires two photo-sensors for each crystal. 

A complementary offline compensation approach is the so-called software compensation, based on shower topology~\cite{calice}. This technique consists of correcting the reconstructed energy on an event-by-event basis, using shower shape or other information, without the use of dual-readout for an intermediate $f_{EM}$ estimation.
It has been shown that software compensation in a Cherenkov-based HCAL with 3D readout can reach the performance of a dual-readout calorimeter~\cite{nural}, using Graph Neural Networks.

Crystal calorimeters present a particularly challenging case. Their excellent electromagnetic performance comes at the price of a highly non-compensating response, which risks spoiling the combined ECAL+HCAL hadron energy resolution, if an HCAL with an excellent resolution is put downstream.
However, modern calorimeters featuring fine longitudinal and transverse segmentation contain detailed information about shower development that can potentially be exploited to recover part of the lost information, and achieve a remarkable performance also in the hadronic energy resolution.

Importantly, non-compensation affects the ECAL and HCAL responses separately, since the electromagnetic fractions in the two detectors are largely independent. In particular, the number of $\pi^0$ produced in the HCAL, which determines its $f_{\mathrm{EM}}$, is independent of the value of $f_{\mathrm{EM}}$ in the ECAL, as discussed in~\cite{lucchini}. Therefore, software compensation could be applied to the ECAL and combined with dual-readout compensation in the HCAL, or vice versa. The need for non-compensation corrections that include the correlations between ECAL and HCAL is excluded in~\cite{lucchini}, but this aspect could be studied in detail with a full detector simulation of ECAL and HCAL, which is beyond the scope of this paper.

In this work, software compensation techniques are investigated for CRILIN~\cite{crilin0, crilin1}, a crystal ECAL based on a semi-homogeneous concept, featuring longitudinally segmented Cherenkov PbF$_2$ crystal matrices and silicon photomultiplier (SiPM) readout. Two approaches were investigated, the first being simple one-dimensional corrections based on shower-shape observables, such as cluster RMS width and longitudinal cluster center-of-gravity. The second approach exploits the full three-dimensional shower topology using a Graph Neural Network, taking as inputs the calorimetric hits and yielding the corrected total reconstructed energy in CRILIN as an output.
In both cases, it is shown that the hadron energy resolution of a combined ECAL+HCAL system significantly improves with respect to non-compensating energy reconstruction.
In Appendix~\ref{app:appendice-dr}, a comparison between the software compensation approach and dual-readout has been performed, showing comparable performances for the two techniques, within the assumptions and limitations of the present study.

\section{CRILIN calorimeter}
The CRILIN calorimeter is an innovative semi-homogeneous electromagnetic calorimeter concept developed for future Muon Collider experiments, which can also be
evaluated as a potential solution for future $e^+e^-$ colliders. Its design 
addresses one of the main challenges of the Muon Collider environment: the large beam-induced background (BIB) generated by muon decays in the accelerator ring. Unlike 
conventional sampling calorimeters, CRILIN combines the excellent energy resolution of homogeneous crystal calorimeters with longitudinal segmentation and high granularity, allowing 
efficient discrimination between physics signals and background particles. 
Overall, the CRILIN concept demonstrates a compelling combination of excellent timing performance, satisfactory energy resolution, radiation tolerance, and cost-effectiveness compared with tungsten--silicon sampling calorimeters, suggested for the same kind of applications, resulting in lower detector complexity and cost~\cite{crilin0}. Despite being originally designed for the detector of a Muon Collider, these requirements are such that CRILIN can also be employed in an FCC-ee detector, if the additional requirements on jet energy resolution are met.

The detector consists of multiple layers of 4~cm long high-density PbF$_2$ Cherenkov crystals read out and interleaved
by UV-extended SiPMs, with a cell size of $1.3 \times 1.3~\mathrm{cm}^2$.

Beam tests performed with the Proto-1 prototype, consisting of two layers of $3\times3$ crystal matrices, demonstrated excellent timing capabilities. Measurements with 120~GeV electron beams at CERN achieved time resolutions below 40~ps for energy deposits exceeding 1~GeV. Furthermore, inter-layer timing measurements yielded a resolution of approximately 30~ps, satisfying the stringent requirements for efficient rejection of beam-induced background~\cite{crilin1,crilin2}.
A full-containment CRILIN module was tested during the summer of 2026 at the CERN SPS H2 beamline with electron beams. The measured energy resolution is characterized by a constant term of approximately 0.23\%, a total noise contribution ranging from $80$ to 200~MeV (depending on beam energy), and a stochastic term of approximately $6.5\%/\sqrt{E[\mathrm{GeV}]}$~\cite{h2june}. This performance lies between the "aggressive" ($3\%/\sqrt{E[\mathrm{GeV}]}$) and the "conservative" ($10\%/\sqrt{E[\mathrm{GeV}]}$) target resolutions for an FCC-ee ECAL~\cite{fcc-ee-vol1}. The measured resolution is reproduced with good agreement through a Monte Carlo Geant4~\cite{geant4} simulation that includes the dominant experimental effects. The CRILIN module was subsequently tested at the CERN PS with negative pion beams in a combined beam test together with the MPGD-HCAL setup~\cite{2024mpgdhcal}.

\section{Geometry and simulation}

\subsection{Detector geometry}

The simulated detector consists of a CRILIN module composed of five longitudinal layers of PbF$_2$ crystals. Each layer contains a $7\times7$ matrix of crystals with dimensions $1.3\times1.3\times4.0~\mathrm{cm}^3$, together with passive elements, corresponding to the crystal wrapping material (Mylar), the alveolar aluminum matrix holding the crystals in place (wall thickness of 200~$\mu$m), and SiPMs and PCBs between layers (2~mm thickness). All geometrical parameters are fully consistent with those of the full-containment module discussed above.
The total calorimeter depth is approximately 20~cm, corresponding to about $21.3~X_0$ and $0.90~\lambda_I$, where the radiation length and nuclear interaction length of PbF$_2$ are $X_0 = 0.9369~\mathrm{cm}$ and $\lambda_I = 22.1~\mathrm{cm}$.
A virtual detector (VD) is placed immediately downstream of CRILIN and is used to estimate the true hadronic energy escaping the crystal calorimeter on the downstream side, $E_{VD}$, working effectively as a truth-level HCAL. In this way, with a known beam energy $E_{beam}$, the true energy deposited in CRILIN can be estimated as $E_{true} = E_{beam} - E_{VD}$.

\begin{figure}[!htbp]
\centering
\includegraphics[height=5cm]{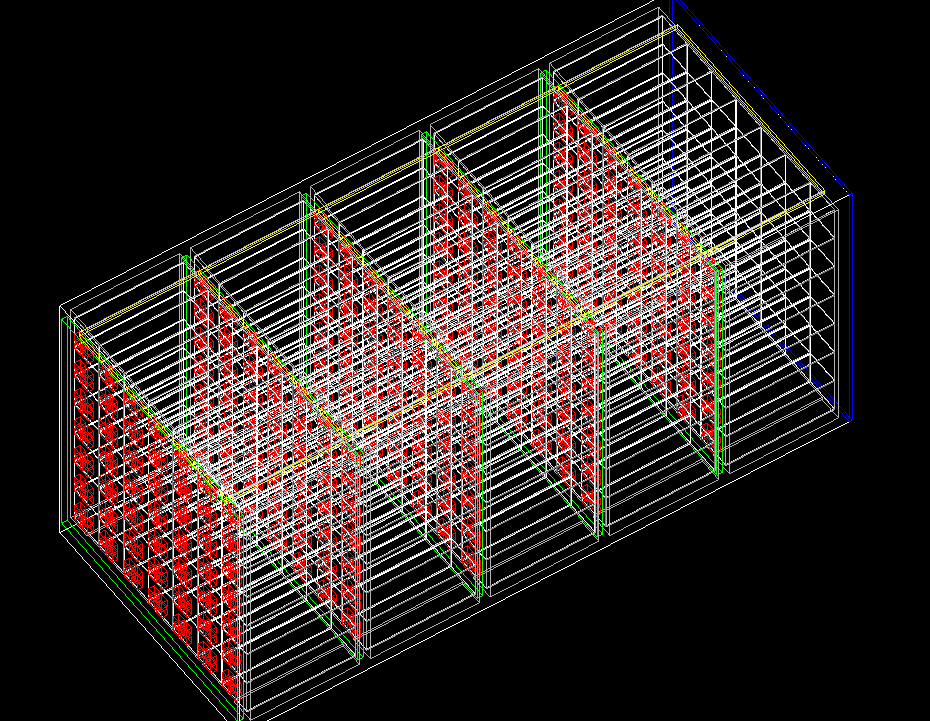}
\hspace{3mm}
\includegraphics[height=5cm]{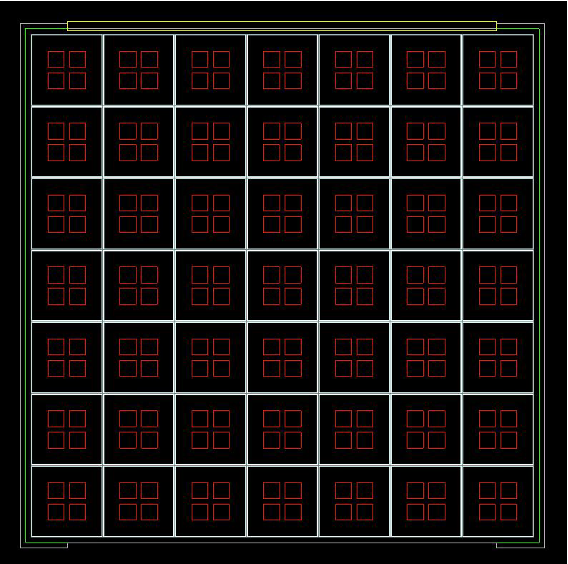}
\caption{Lateral (left) and frontal (right) views of the Geant4-simulated CRILIN module, including the complete geometry and mechanical structure of the final detector. The edges of the aluminum matrix, SiPM sensors, SiPM PCB boards, Kapton readout strips, and virtual detector are shown, respectively, in white, red, green, yellow, and blue.}
\label{fig:geometry}
\end{figure}


\subsection{Simulation and readout}

The detector response is simulated using Geant4 with the QGSP\_BERT physics list~\cite{geant4}. The choice of Geant4 physics list may affect the quantitative results; however, a detailed evaluation of this dependence is beyond the scope of the present study. Single charged pions were generated with energies between 15 and 100~GeV. Rather than performing a full optical simulation, the Cherenkov response is estimated using the Frank--Tamm formula~\cite{frank-tamm}:
\begin{equation}
\frac{\mathrm{d}^2 N}{\mathrm{d}x\,\mathrm{d}\lambda}
=
\frac{2\pi\alpha}{\lambda^2}
\left(
1-\frac{1}{\beta^2 n^2(\lambda)}
\right),
\end{equation}

where $N$ is the number of emitted Cherenkov photons, $x$ is the particle path length in the medium, $\lambda$ is the vacuum wavelength of the emitted radiation, $\alpha$ is the fine-structure constant, $\beta$ is the particle speed divided by the speed of light, and $n(\lambda)$ is the refractive index of the medium.

\begin{figure}[!htbp]
\centering
\includegraphics[width=0.45\linewidth]{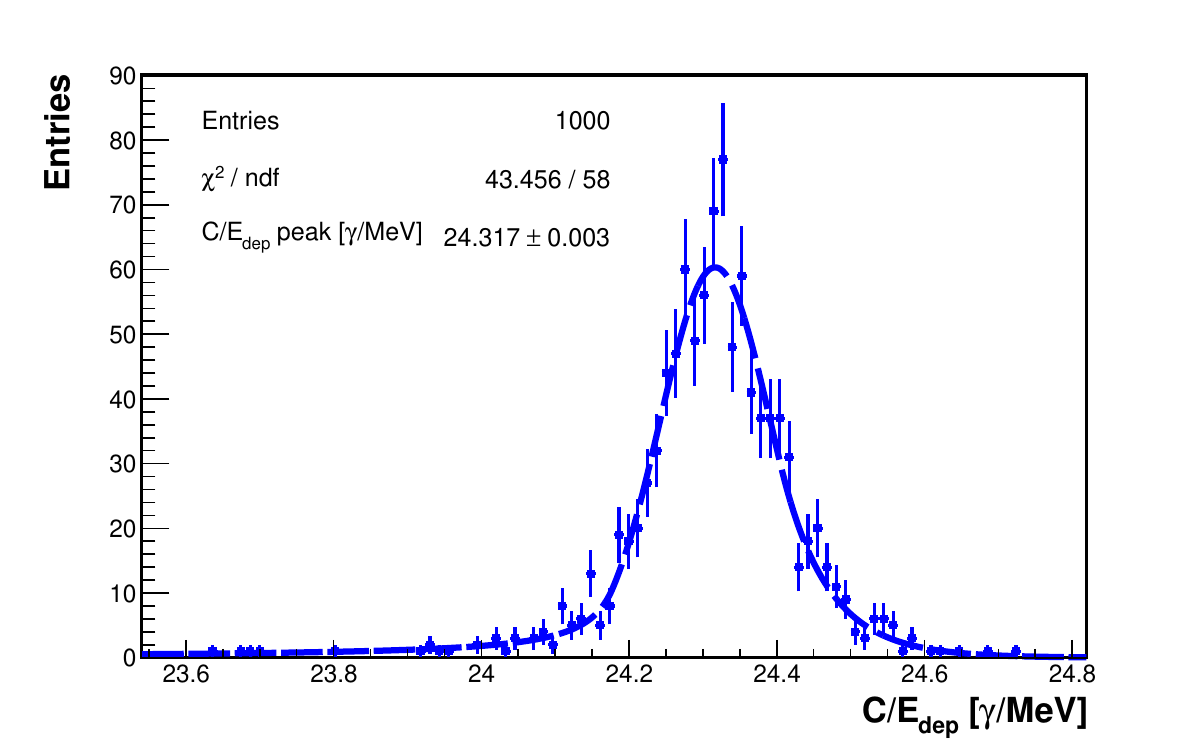}
\includegraphics[width=0.45\linewidth]{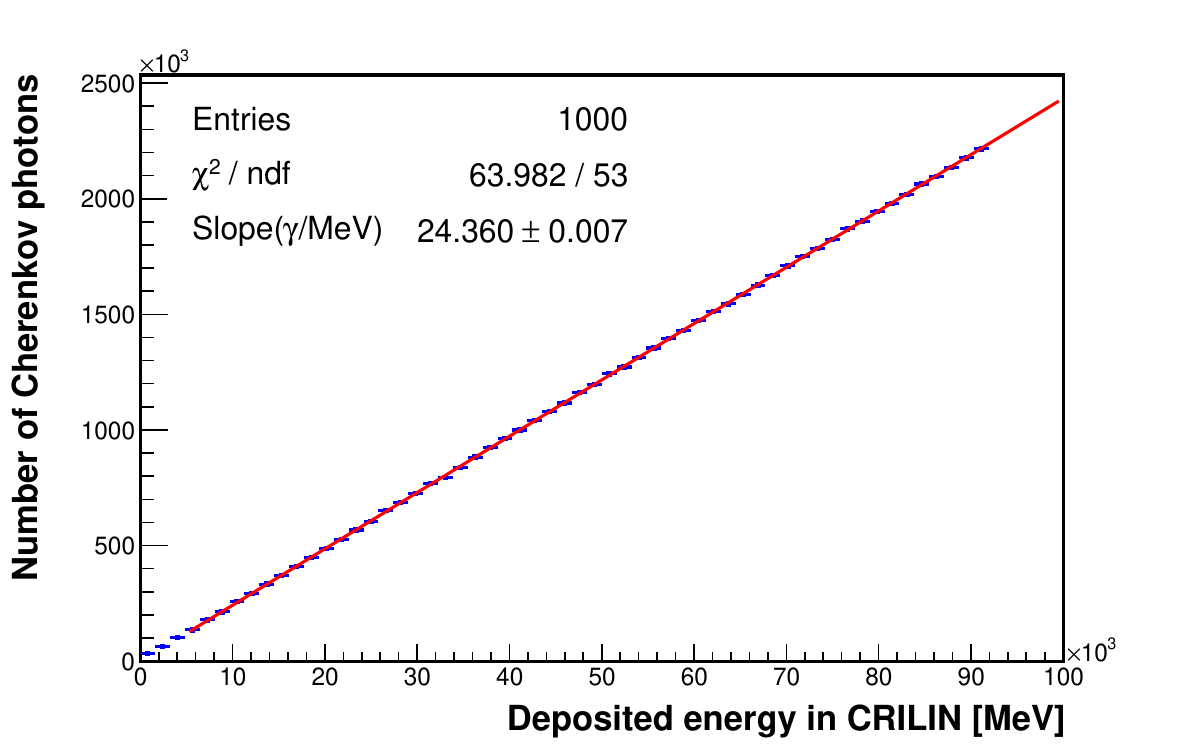}
\caption{Left: distribution of the ratio between number of emitted Cherenkov photons and deposited energy in the CRILIN module, $C/E_{dep}$, fitted with a Double Crystal Ball function~\cite{dcb}. Right: profile histogram of the number of emitted Cherenkov photons as a function of deposited energy in the CRILIN module, fitted with a linear function.
Both figures are based on a simulation with a uniform electron beam energy between 1 and 100~GeV.}
\label{fig:electrons}
\end{figure}

Therefore, the number of emitted Cherenkov photons is computed in the 350--550 nm wavelength range, selected as an approximation of the effective spectral sensitivity of the SiPM-based readout, from the charged-particle track lengths, velocities, and charges recorded in the Geant4 simulation. To convert the number of emitted photons into an energy value, a dedicated simulation is performed with electrons between 1 and 100~GeV, and the Cherenkov light response ($C$) is compared to the total energy deposited in the calorimeter ($E_{dep}$).
The results are shown in Figure~\ref{fig:electrons}, using both the distribution of the ratio $C/E_{dep}$ and a linear fit to the $C$ vs. $E_{dep}$ distribution, yielding a conversion factor of approximately $C/E_{dep} = 24.32~\gamma$ /MeV. The difference in the $C/E_{dep}$ ratios found with the two approaches is approximately 0.16\%, which is sufficiently small for the purposes of this paper. This ratio is dependent on the physics list at the level of at least 20\%~\cite{h2june}, but this has no significant effect on the conclusions of the analysis presented in this paper.

The photon-to-energy conversion derived from the electron sample provides an effective calibration of the average Cherenkov response, but it does not account for the photosensor photon-detection efficiency or the light-collection efficiency. It therefore cannot be used to estimate the calorimeter energy resolution arising from photo-statistics.
For this reason, a global factor comprising both efficiencies is evaluated starting from the measured photoelectron (p.e.) light yield of the crystal–photosensor system. In this work, a conservative light yield value of \(0.2~\mathrm{p.e./MeV}\) is assumed, which is lower than the value measured with previous CRILIN prototypes~\cite{ieee-CRILIN}, and with the full-containment module~\cite{h2june}, i.e. 0.54 p.e./MeV.

Poisson fluctuations corresponding to this light yield are applied on an event-by-event basis, and crystal hits below \(40~\mathrm{MeV}\) are discarded to emulate the hit-reconstruction threshold used in realistic detectors to suppress electronic noise. Unless otherwise stated, all hits used throughout this paper are preprocessed with this Poisson smearing and threshold requirement.

This conversion factor has been cross-checked using a 5~GeV muon sample, for which the ratio between the number of Cherenkov photons and the deposited energy, \(E_{\mathrm{dep}}\), is found not to be constant as a function of \(E_{\mathrm{dep}}\), as shown in Appendix~\ref{app:appendice-muoni}, Figure~\ref{fig:mu-nonline}. This effect has been understood and correctly taken into account throughout the paper, as described in Appendix~\ref{app:appendice-muoni}. In particular, at the minimum ionizing energy deposit, the difference between muons and showering-electrons conversion factor is around 50\%. 


\section{Software compensation using shower observables}

\begin{figure}[!htbp]
\centering
\includegraphics[width=0.32\linewidth]{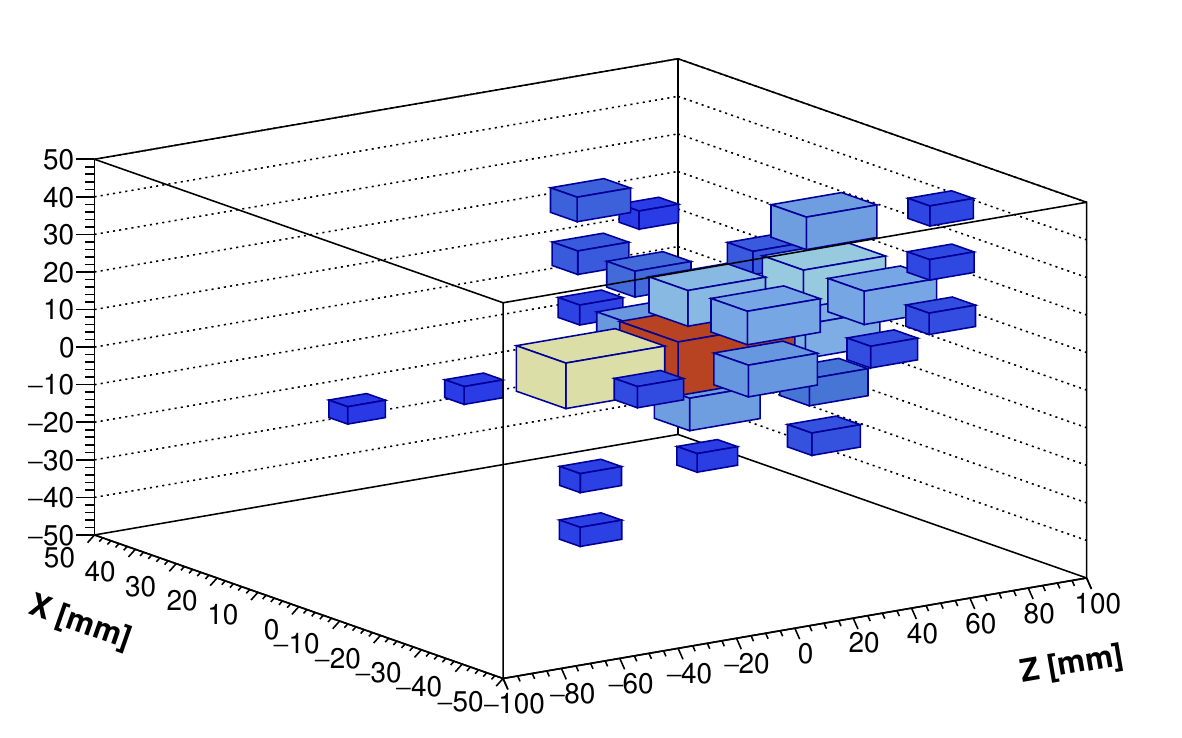}
\includegraphics[width=0.32\linewidth]{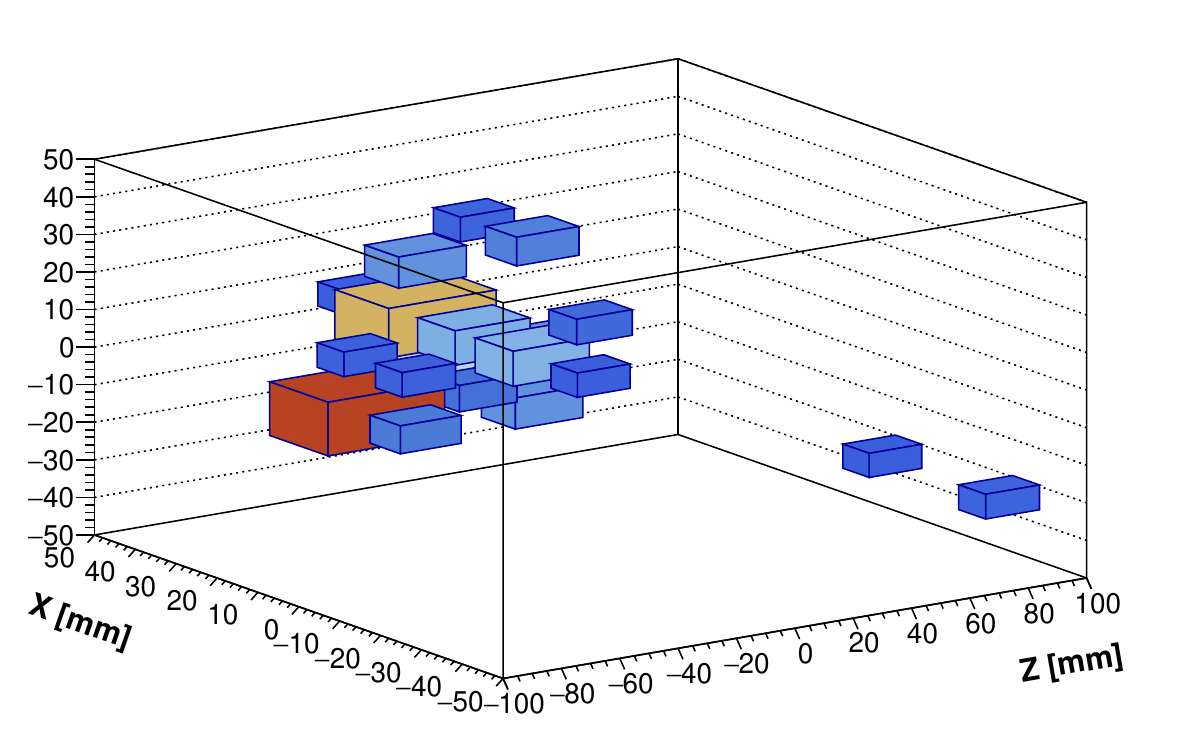}
\includegraphics[width=0.32\linewidth]{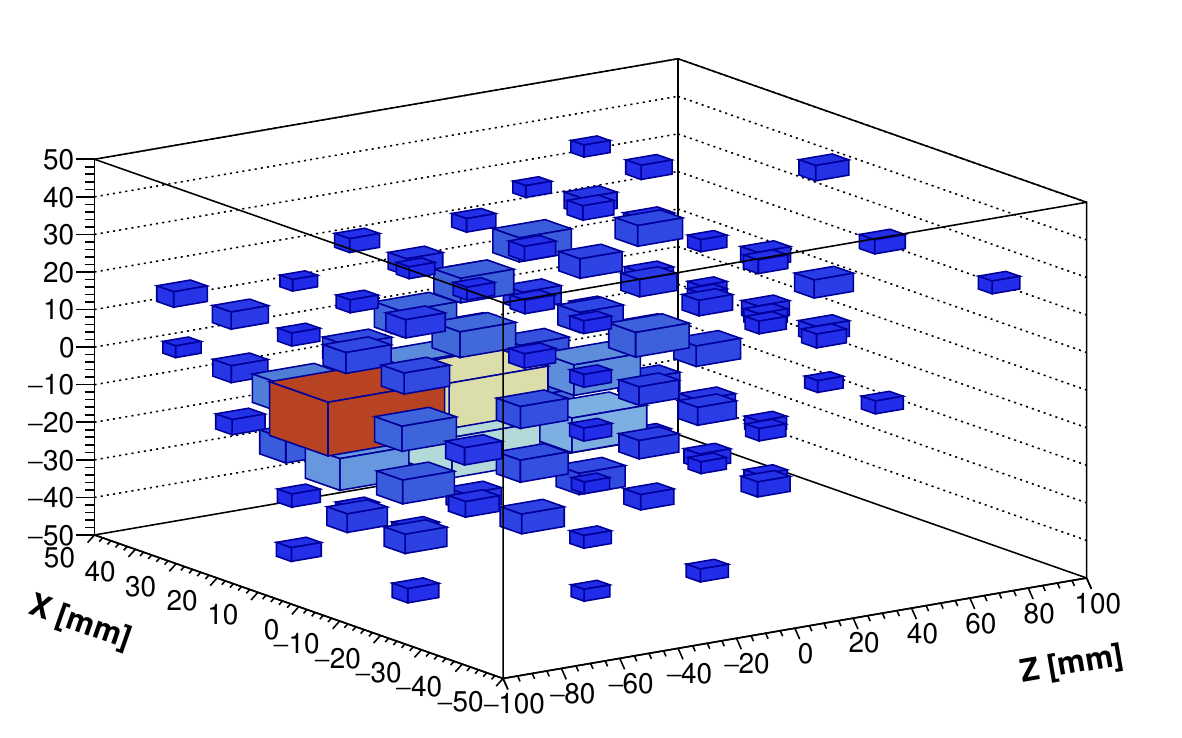}
\caption{Three event displays for simulated pion showers in the CRILIN calorimeter.}
\label{fig:ev}
\end{figure}

Large event-by-event fluctuations are expected in the reconstructed pion energy, originating from variations of the electromagnetic fraction of hadronic showers. Significantly different shower topologies can readily be observed when looking at event displays from different events, as shown in Figure~\ref{fig:ev}.

The first approach investigated in this work consists of exploiting global shower-shape observables in order to evaluate event-by-event energy corrections. This technique is already used in LHC detectors, as detailed in~\cite{atlas}.


\subsection{Transverse shower RMS}
\label{rms}
The transverse RMS of the shower is defined as

\begin{equation}
R_{\mathrm{RMS}}
=
\sqrt{
\frac{
\sum_i E_i |\vec{r}_i-\vec{r}_{\mathrm{cog}}|^2
}
{
\sum_i E_i
}
}.
\end{equation}
where $\vec{r}_{cog}$ is the shower center-of-gravity (cog).
The primary quantity used throughout this work to derive energy corrections is the ratio $E_{reco} / E_{true}$ between the reconstructed energy in CRILIN, i.e. the simple sum of the hits, and the true energy deposited inside CRILIN, estimated
from the beam and VD energy as described above.

\begin{figure}[!htbp]
\centering

\includegraphics[width=0.7\linewidth]{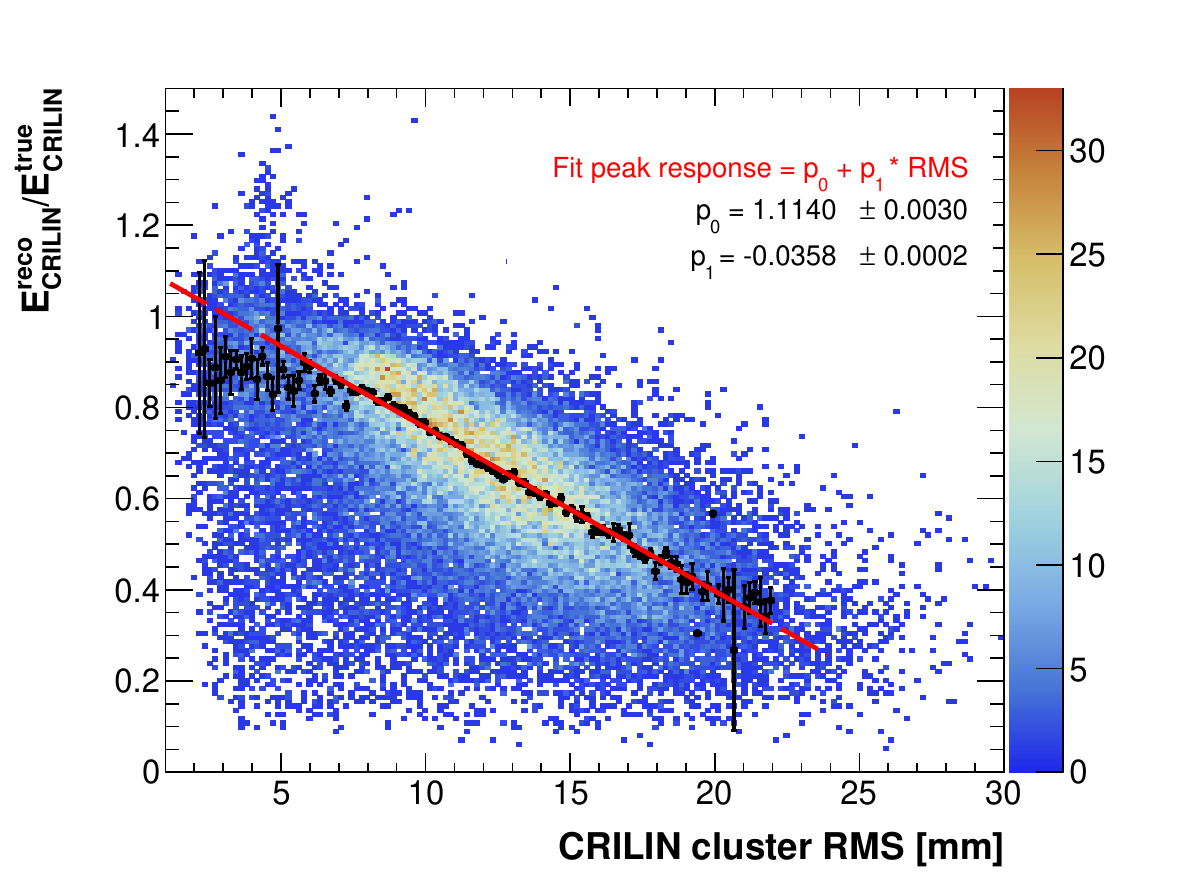}
\caption{Correlation between $E_{reco} / E_{true}$ and shower RMS for a simulated pion sample with uniform energies from 5 to 100~GeV, with a linear fit superimposed.}
\label{fig:rms-overall}

\end{figure}

Events with a larger hadronic deposition tend to exhibit broader shower profiles, leading 
to a correlation between $E_{reco} / E_{true}$ and $R_{\mathrm{RMS}}$, as shown in 
Figure~\ref{fig:rms-overall}.
After binning the events in RMS, the corresponding peaks in the projected
$E_{reco} / E_{true}$ distribution have been fitted with a linear fit, superimposed in the Figure. 
It should be noted that 
the tail of the RMS distribution is around 20~mm, which means that in crystal 
calorimeters the cell areas should be small enough in order to resolve these 
correlations.

A linear function is fitted in intervals of beam energy with 5~GeV width according to the formula $E_{reco} / E_{true} = p_0+p_1 R_{\mathrm{RMS}}$, and the corresponding 
correction is applied as: \begin{equation} E_{\mathrm{corr}} = 
\frac{E_{\mathrm{reco}}}
     {p_0+p_1 R_{\mathrm{RMS}}}
\end{equation}
with parameters $p_0, p_1$ depending on the energy interval considered.

In order to estimate the impact of this correction on the resolution of a combined calorimetric system with both an ECAL and a HCAL section, the virtual detector energies have been smeared with a Gaussian fluctuation corresponding to a relative resolution in the range between $25\%/\sqrt{E[\mathrm{GeV}]}$ and $50\%/\sqrt{E[\mathrm{GeV}]}$, yielding a variable $E_{HCAL}^{25\%/\sqrt{E}}$, mimicking the behavior of a compensating or dual-readout HCAL with energy resolution dominated by the stochastic term and only Gaussian tails~\cite{wigmans-dual-readout}. In this case, the energy that enters the resolution formula $S/\sqrt{E[\mathrm{GeV}]}$ is the energy released in the HCAL, $E_{HCAL}$. In the following, for simplicity, the energy is always assumed to be measured in GeV.
As discussed in the following, the combined energy resolution is studied as a function of the assumed HCAL resolution, showing that the value of the HCAL stochastic term has a negligible impact on the effective contribution of CRILIN to the hadron energy resolution.
For the first part of the paper, a value of $25\%/\sqrt{E}$, in the extreme end of the range considered, is employed.

\begin{figure}[!htbp]
\centering
\includegraphics[width=0.49\linewidth]{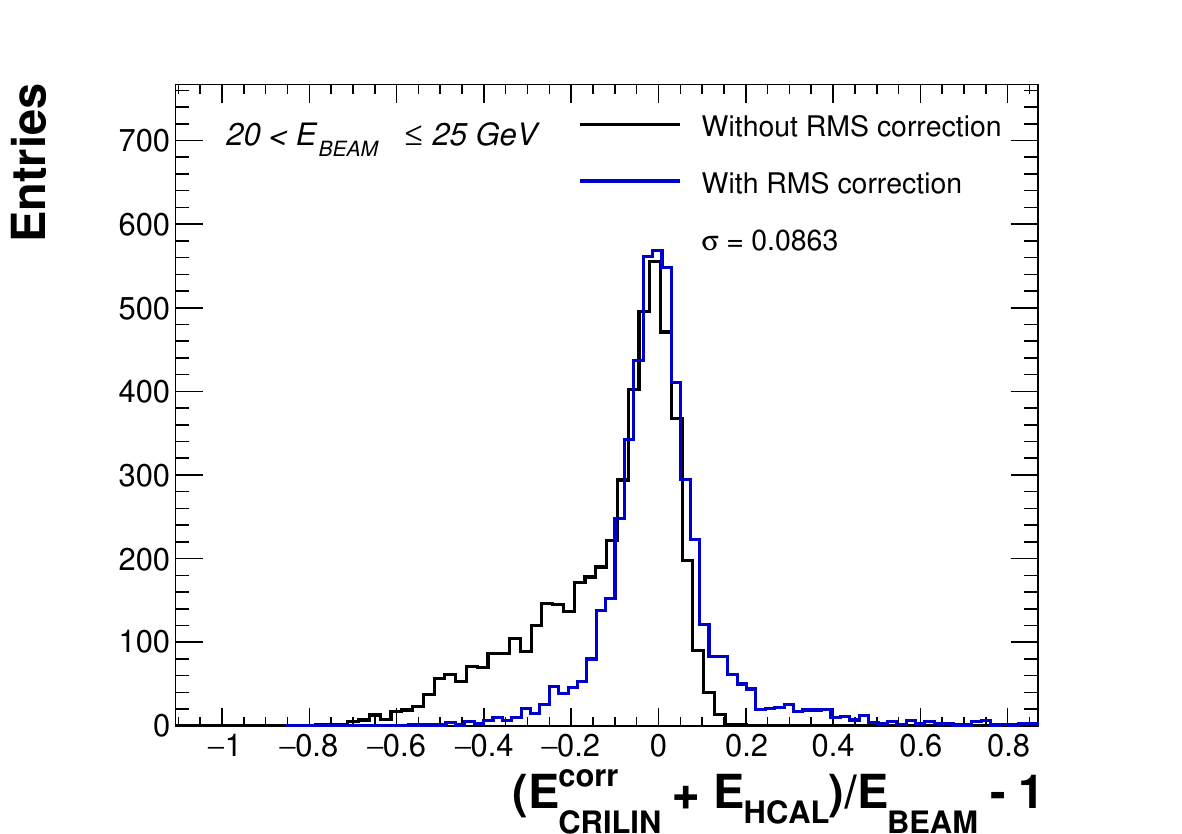}
\hspace{1mm}
\includegraphics[width=0.49\linewidth]{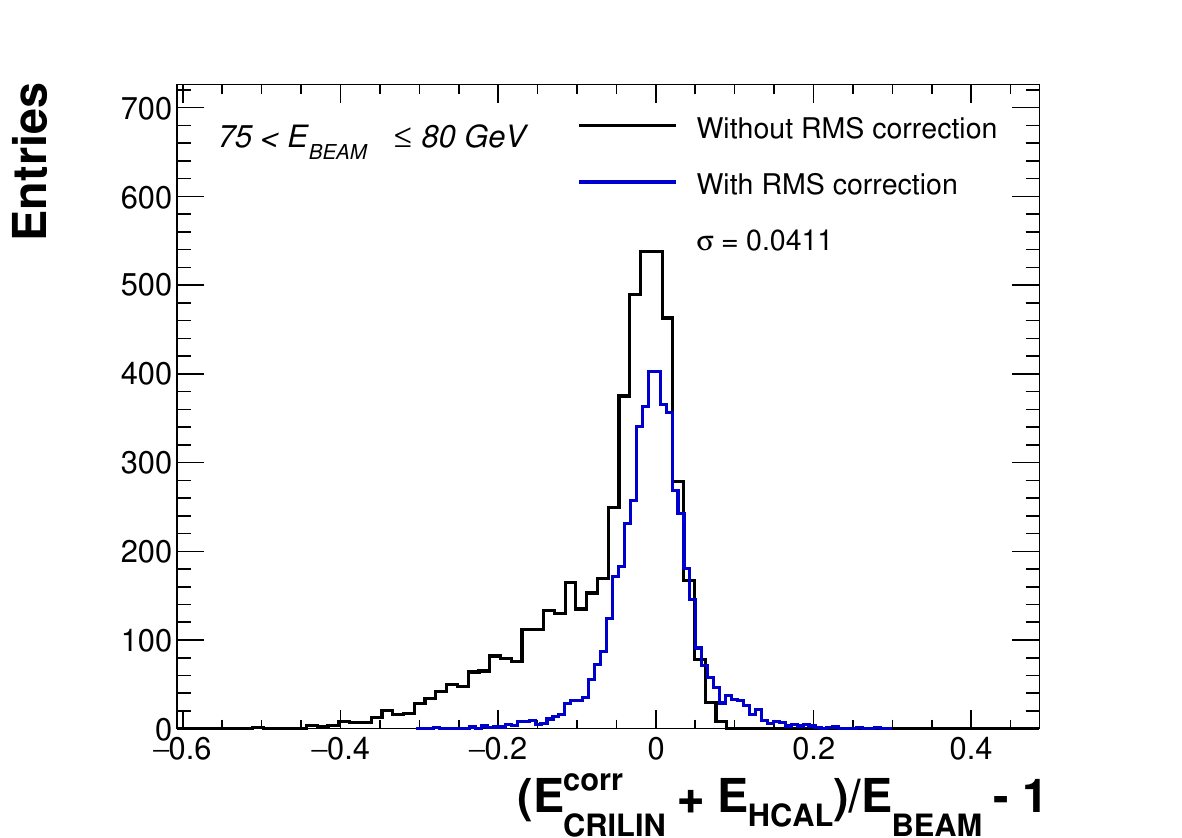}
\caption{Distribution of $(E_{\mathrm{corr}} + E_{HCAL}^{25\%/\sqrt{E}})/E_{beam}$ for a combined CRILIN-HCAL system, with and without the RMS-based correction, using simulated pion samples in two beam energy intervals: 20--25~GeV and 75--80~GeV. The resolutions ($\sigma$) quoted in the plot correspond to the case with corrections applied.}
\label{fig:rms-slices}
\end{figure}

The resolution is therefore evaluated from the distributions of $E_{\mathrm{corr}} + E_{HCAL}^{25\%/\sqrt{E}}$ divided by the beam energy, as shown in Figure~\ref{fig:rms-slices} for the 20--25 and 75--80~GeV beam energy intervals, comparing the distributions before and after the correction.

Throughout the paper, for each energy interval, the resolution is estimated as the half-width of the central 68\% interval, defined between the 16th and 84th percentiles of the distribution, as used in~\cite{cld-ml}.
For a Gaussian distribution, this definition of the resolution is equal to the one estimated using the full-width-half-maximum (FWHM) divided by approximately 2.35, while for distributions with tails larger than Gaussian ones, it is larger than FWHM/2.35, correctly taking into account the effect of the tails.

\begin{figure}[!htbp]
\centering
\includegraphics[width=0.7\linewidth]{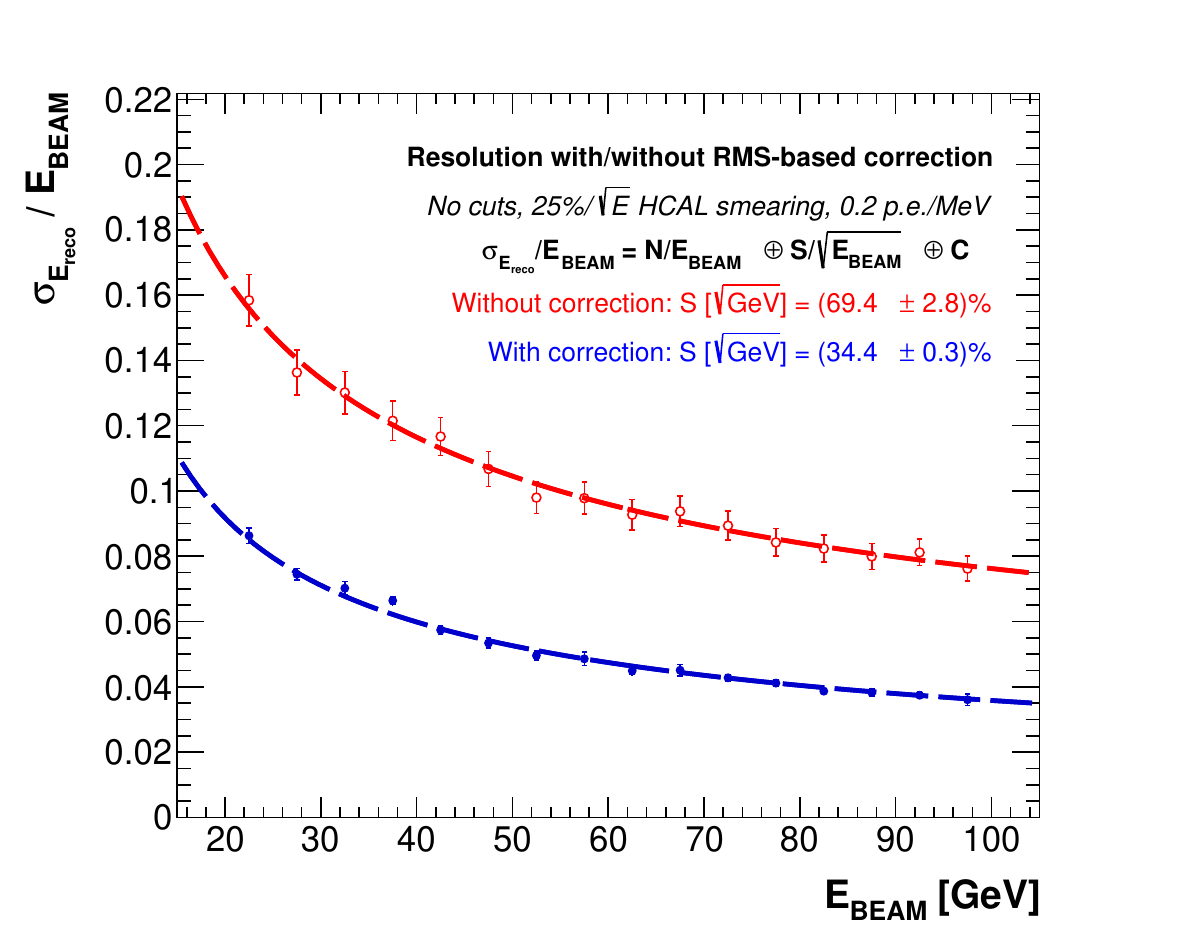}
\caption{Energy resolution for a combined CRILIN and HCAL system, as a function of the beam energy of the simulated pion sample, with and without RMS-based correction applied. The curves are fitted using a three-component model fit, described in the text.}
\label{fig:rms-reso}
\end{figure}

The resolution in the different beam energy intervals is shown in Figure~\ref{fig:rms-reso}, before and after the application of the correction.

The resolution as a function of beam energy is fitted with a three-component model: $\sigma_{E}/E = N / E \, \oplus S/\sqrt{E} \, \oplus C$, where the coefficients $N$, $S$ and $C$ correspond to noise-like, stochastic and constant terms, yielding the values shown in Table~\ref{tab:tab-rms}.
The noise $N$ and constant $C$ terms are more likely to be dominated by instrumental effects in a realistic detector, therefore a more specific focus is given to the stochastic term $S$, which is the only one included within the figures.

In particular, the stochastic term ($S\cdot\sqrt{E}$) is reduced from $(69.4 \pm 2.8)\%$ before the correction to $(34.4 \pm 0.3)\%$ after the correction, showing that this method is effective in recovering an excellent resolution despite the non-compensating nature of the crystal-based calorimeter.

In order to further study the possibilities of this method, a comparison between these results and those obtained with a CRILIN-like scintillating crystal calorimeter, motivated by potential improvements due to the higher light yield, has been carried out in Appendix~\ref{app:appendice-scint}, showing no significant differences.

On the other hand, the physical origin of the correlation between $E_{reco} / E_{true}$ and the RMS is studied in detail in Appendix~\ref{app:appendice-rms}. In particular, the cluster RMS is shown to be strongly correlated with the electromagnetic fraction $f_{EM}$. A simple phenomenological model with two Gaussian shower sub-components reproduces the observed trend and explains why a linear correction provides a reasonable approximation.

\begin{table}[!htbp]
\centering
\begin{tabular}{lcc}
\hline
 & Without correction & With RMS-based correction \\
\hline
$N$ [GeV] & $1.09 \pm 0.03$ & $1.01 \pm 0.02$ \\
$S$ [GeV$^{0.5}$] & $(69.4 \pm 2.8)\%$ & $(34.4 \pm 0.3)\%$ \\
$C$ & $(2.98 \pm 0.05)\%$ & $(0.52 \pm 0.02)\%$ \\
\hline
\end{tabular}
\caption{Values of the coefficients from the three-component resolution fit, in the case of no corrections applied and with RMS-based correction.}
\label{tab:tab-rms}
\end{table}

The noise term of approximately 1~GeV comes from the relatively large thresholds applied to single hits, combined with the fact that Cherenkov light readout collects only energy from relativistic particles. 
Although this term may seem worrying for the electromagnetic performance, which is beyond the scope of this paper, this is not the case. For simulated electron samples with a Cherenkov readout, noise terms were found to be orders of magnitude lower, with the residual contribution in a realistic detector being dominated by instrumental effects. In recent beam test measurements with electrons, the noise term is found to be between 80 and 200~MeV in the 10--120~GeV beam energy range~\cite{h2june}.

\subsection{Longitudinal center-of-gravity}

\label{z}
A complementary observable is the energy-weighted longitudinal center-of-gravity,

\begin{equation}
Z_{\mathrm{cog}}
=
\frac{
\sum_i E_i z_i
}{
\sum_i E_i
}.
\end{equation}

Since the electromagnetic component of the shower develops earlier than the hadronic component, $Z_{\mathrm{cog}}$ provides additional sensitivity to fluctuations of the electromagnetic fraction, but its effectiveness is reduced because the shower starting point fluctuates according to an exponential distribution with an average of around 1 nuclear interaction length, which is comparable to the full longitudinal size of the CRILIN calorimeter.

\begin{figure}[!htbp]
\centering
\includegraphics[width=0.48\linewidth]{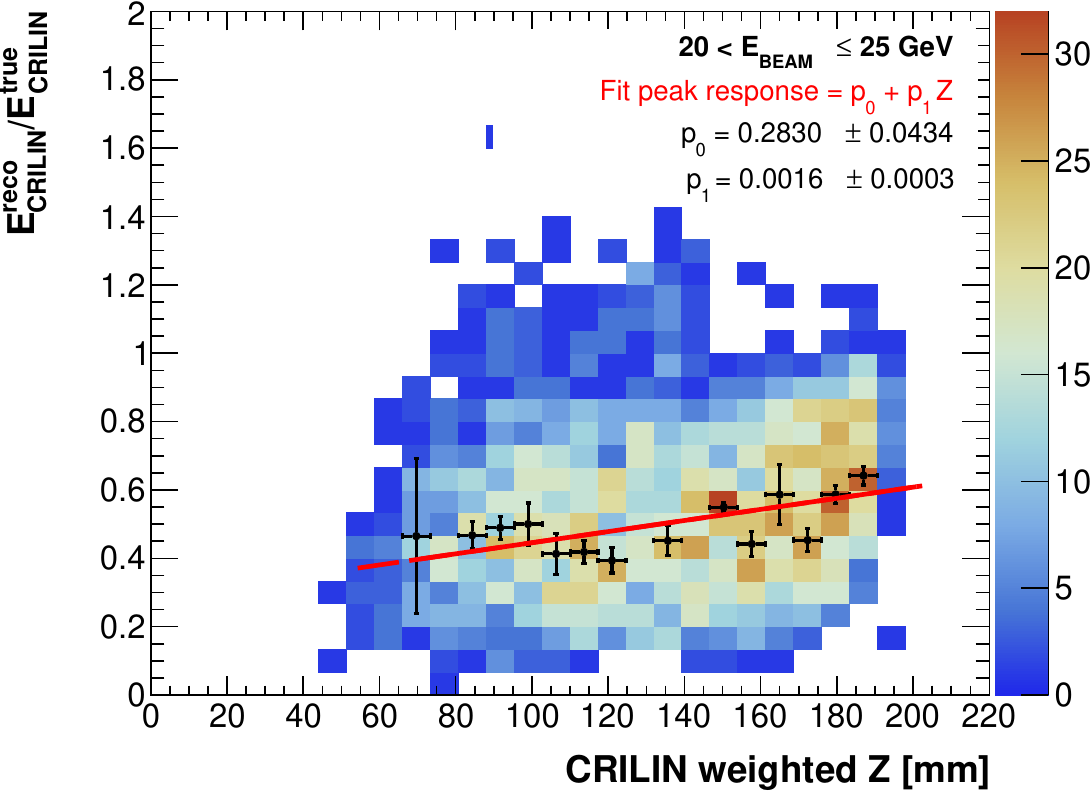}
\hspace{4mm}
\includegraphics[width=0.48\linewidth]{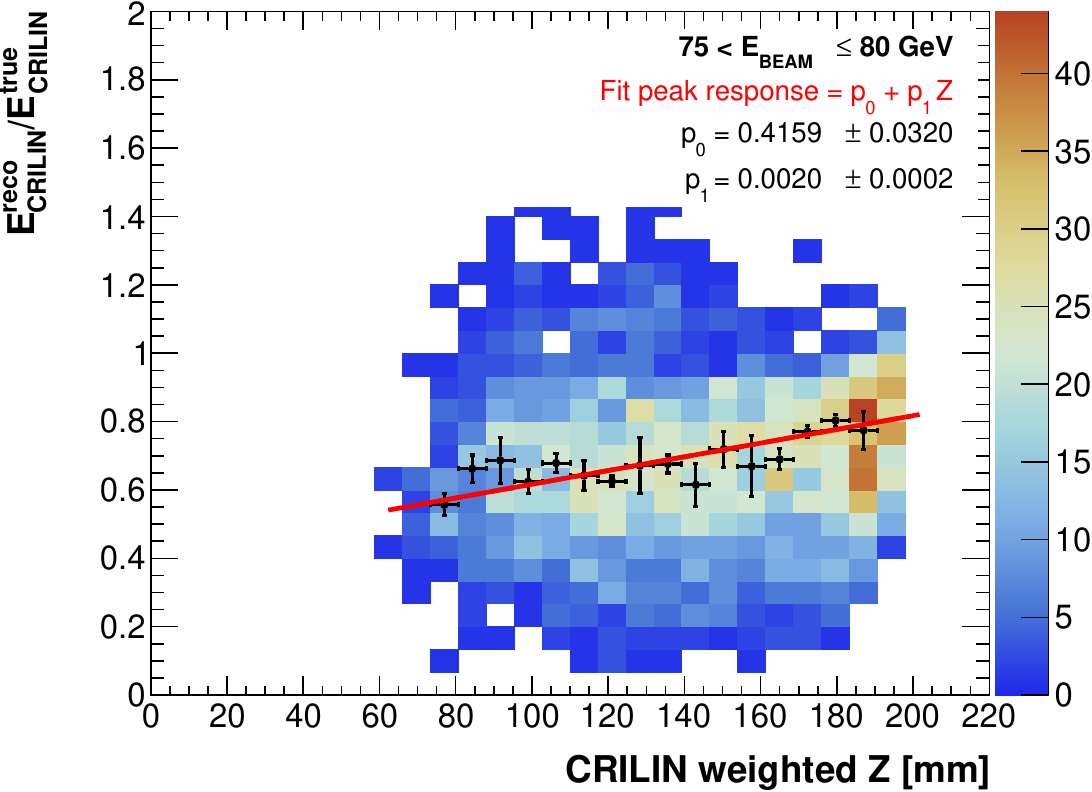}
\caption{Correlation between reconstructed energy and longitudinal center-of-gravity for a simulated pion sample in two beam energy intervals, with a linear fit superimposed.}
\label{fig:z-corr}
\end{figure}

The correlation between $E_{reco}/E_{true}$ and the longitudinal center-of-gravity is shown in Figure~\ref{fig:z-corr}, with linear fits superimposed, for two beam energy intervals, i.e. 20--25~GeV and 75--80~GeV.
The linear fit parameters are used to correct the reconstructed energy on an event-by-event basis as a function of $Z_{\mathrm{cog}}$, yielding a variable $E_{corr}$, in a manner similar to the RMS-based correction.

\begin{figure}[!htbp]
\centering
\includegraphics[width=0.7\linewidth]{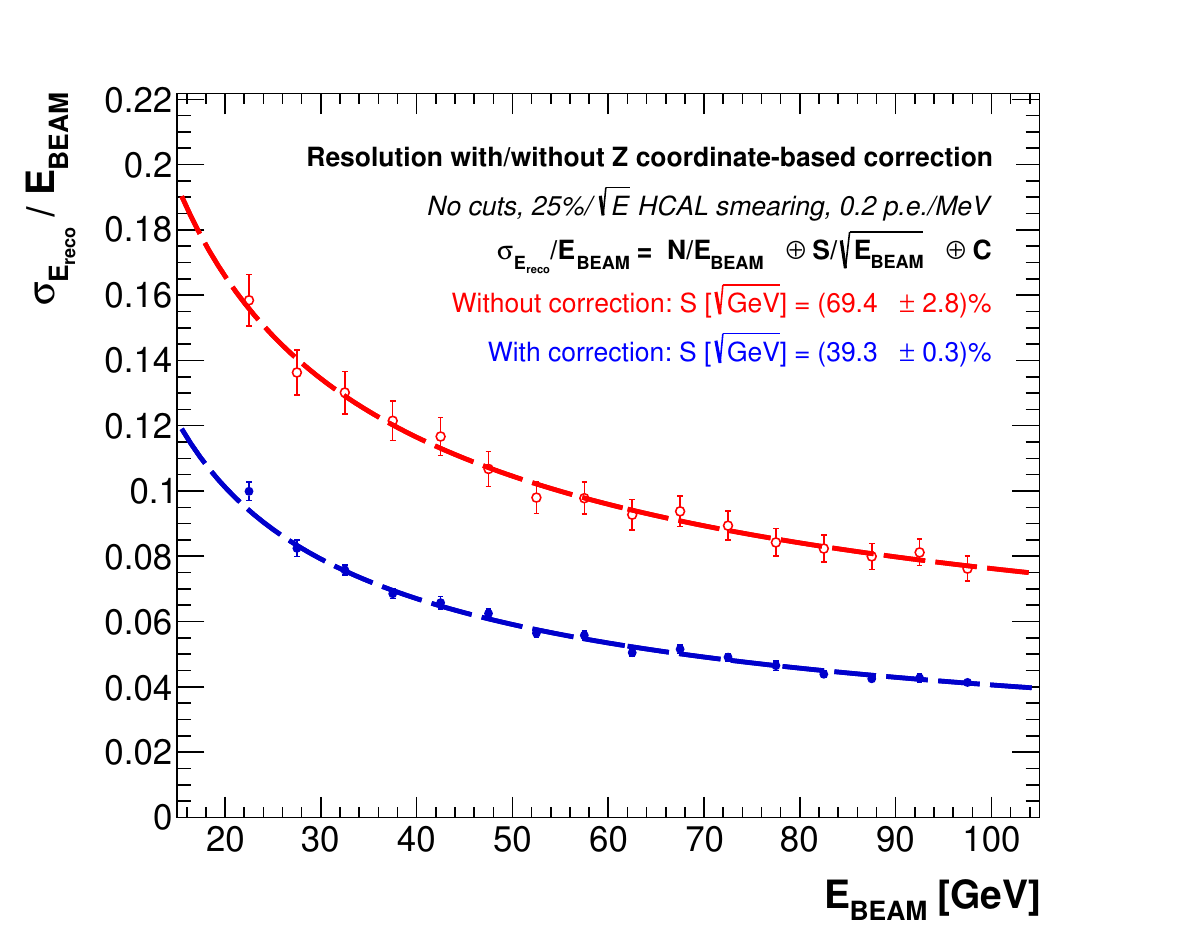}
\caption{Energy resolution for a simulated pion sample as a function of the beam energy, with and without the longitudinal center-of-gravity--based correction. The curves are fitted using a three-component model fit, described in the text.}
\label{fig:z-reso}
\end{figure}

The energy resolution in different beam energy intervals, evaluated from the distributions of $(E_{corr} + E_{HCAL}^{25\%/\sqrt{E}})/E_{beam}$, with similar HCAL smearing as in the RMS case, is shown in Figure~\ref{fig:z-reso} as a function of beam energy. The same three-component fit employed above has been used, yielding a stochastic term ($S\cdot\sqrt{E}$) of $(39.3 \pm 0.3)$\% after the correction, which is 14\% larger 
than the value found for the RMS-based correction.

The possibility of applying an additional Z-based correction after the RMS-based ones has been investigated in Appendix~\ref{app:appendice-z}, showing that there is no significant gain with respect to the pure RMS-based corrections.

\section{ParticleNet training and results}
In this study, the one-dimensional corrections (RMS- and Z-based) are derived in narrow beam energy intervals, using energy-dependent coefficients. However, these global shower observables exploit only a small fraction of the available information: CRILIN provides detailed three-dimensional hit-level measurements that can be processed using Graph Neural Networks. The longitudinal information provided by CRILIN is particularly important, as pion showers tend to start closer to the downstream end of the ECAL. In principle, timing information could also be employed, but this goes beyond the scope of this paper.
For these two reasons, a more complex approach involving neural networks has been investigated, employing only information available at the reconstruction stage.

The input graph is built from reconstructed calorimeter hits. Each hit is represented by four features, i.e. the three-dimensional position and the energy.
The network architecture is based on ParticleNet, a Graph Neural Network (GNN), developed for application in high-energy particle physics~\cite{particlenet}, and consists of three EdgeConv blocks with $k=50$ nearest neighbors and feature dimensions of 32, 32 and 64 channels, respectively, followed by a global average pooling layer.

The output of the network is the corrected reconstructed energy $E_{GNN}$, obtained from the ParticleNet block using a series of three dense layers, with 128, 64 and 1 neurons respectively, ReLU activation, and interleaved with dropout layers.
The network target is the true energy deposited inside CRILIN, $E_{\mathrm{true}}^{\mathrm{CRILIN}}
=
E_{\mathrm{beam}}
-
E_{\mathrm{VD}}$.

Training is performed using the Adam optimizer with learning rate equal to $10^{-4}$, batch size 64 and mean-squared-error loss. 
The network is trained for 60 epochs on a Tesla T4 GPU, using 20\% of the dataset for validation, starting from the same dataset used for RMS and Z-based corrections. 
The test dataset is an independent simulated sample with a different random seed.

\begin{figure}[!htbp]
\centering
\includegraphics[width=0.475\linewidth]{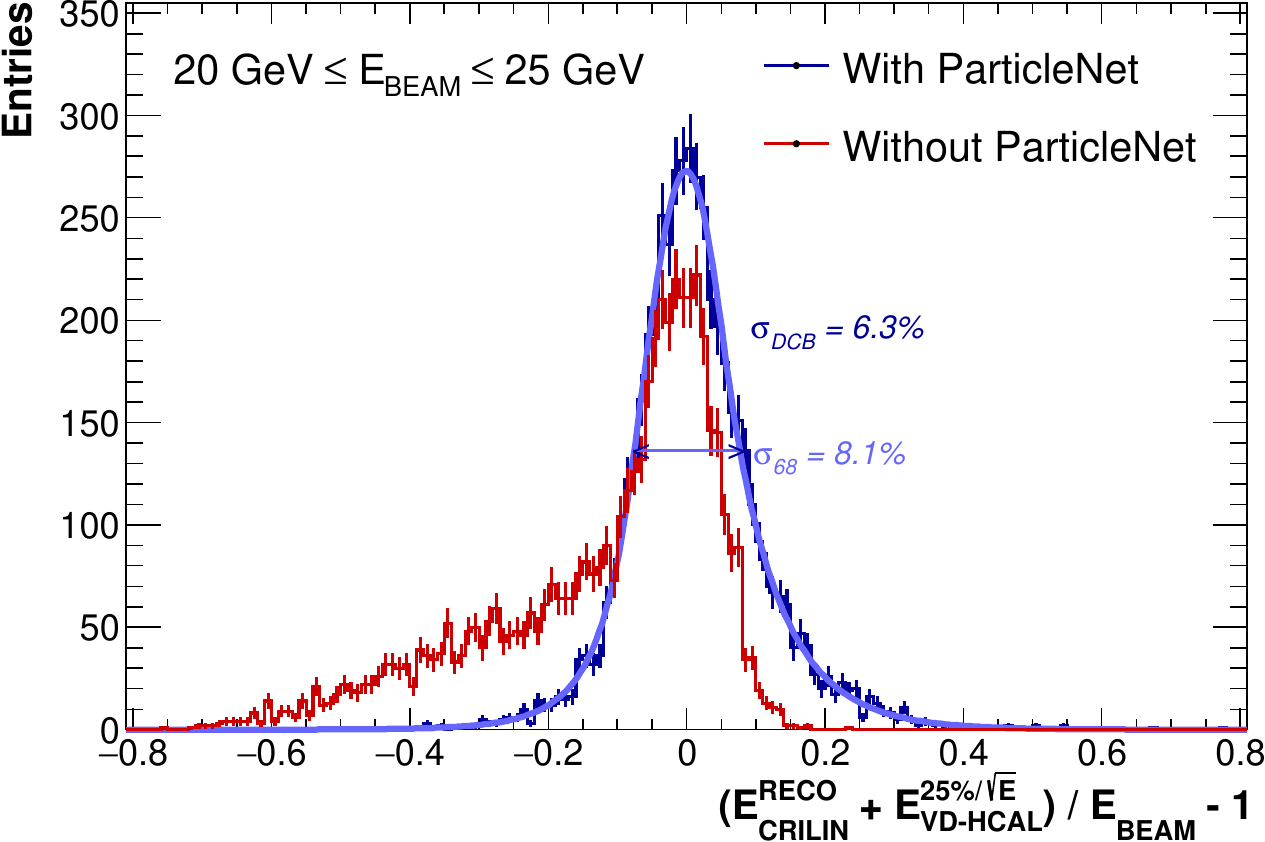}
\hspace{4mm}
\includegraphics[width=0.475\linewidth]{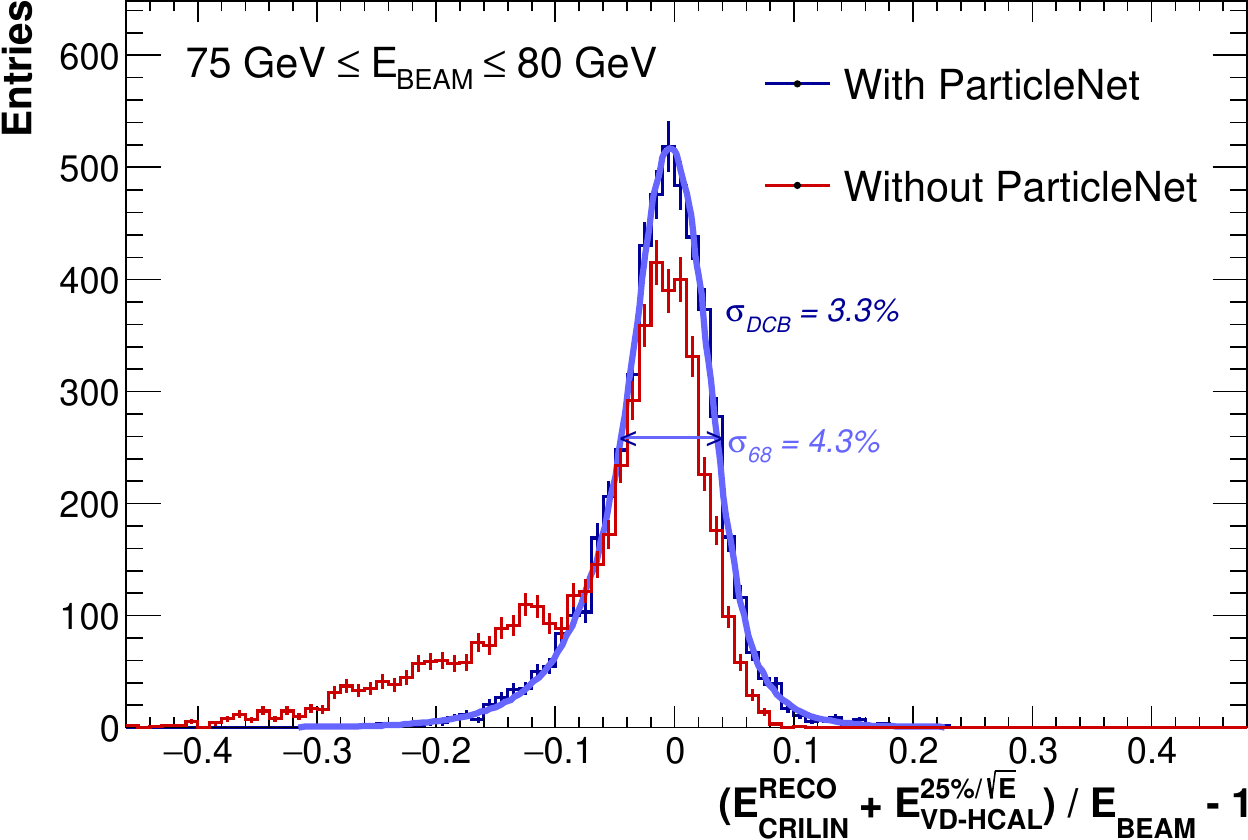}
\caption{Distributions of $E_{reco}/E_{beam} - 1$ for two beam energy intervals with a simulated pion sample, i.e. 20--25~GeV and 75--80~GeV, with and without the GNN-based correction. The corrected distribution was fitted with a Double Crystal Ball function. The resolution ($\sigma_{68}$) corresponding to the central 68\% quantile interval is also evaluated.}
\label{fig:rescomparison}
\end{figure}

Figure~\ref{fig:rescomparison} shows two distributions of $(E_{reco}/E_{beam} - 1) $, which correspond to the relative fluctuations of reconstructed energy $E_{reco} = E_{GNN} + E_{HCAL}^{25\%/\sqrt{E}}$ with and without the GNN-based correction, in two beam energy intervals, i.e. 20--25~GeV and 75--80~GeV, on the test sample. The distributions are fitted with a Double Crystal Ball function, and the resolution corresponding to the central 68\% quantile interval (as throughout the rest of the paper) is evaluated.

\begin{figure}[!htbp]
\centering
\includegraphics[width=0.7\linewidth]{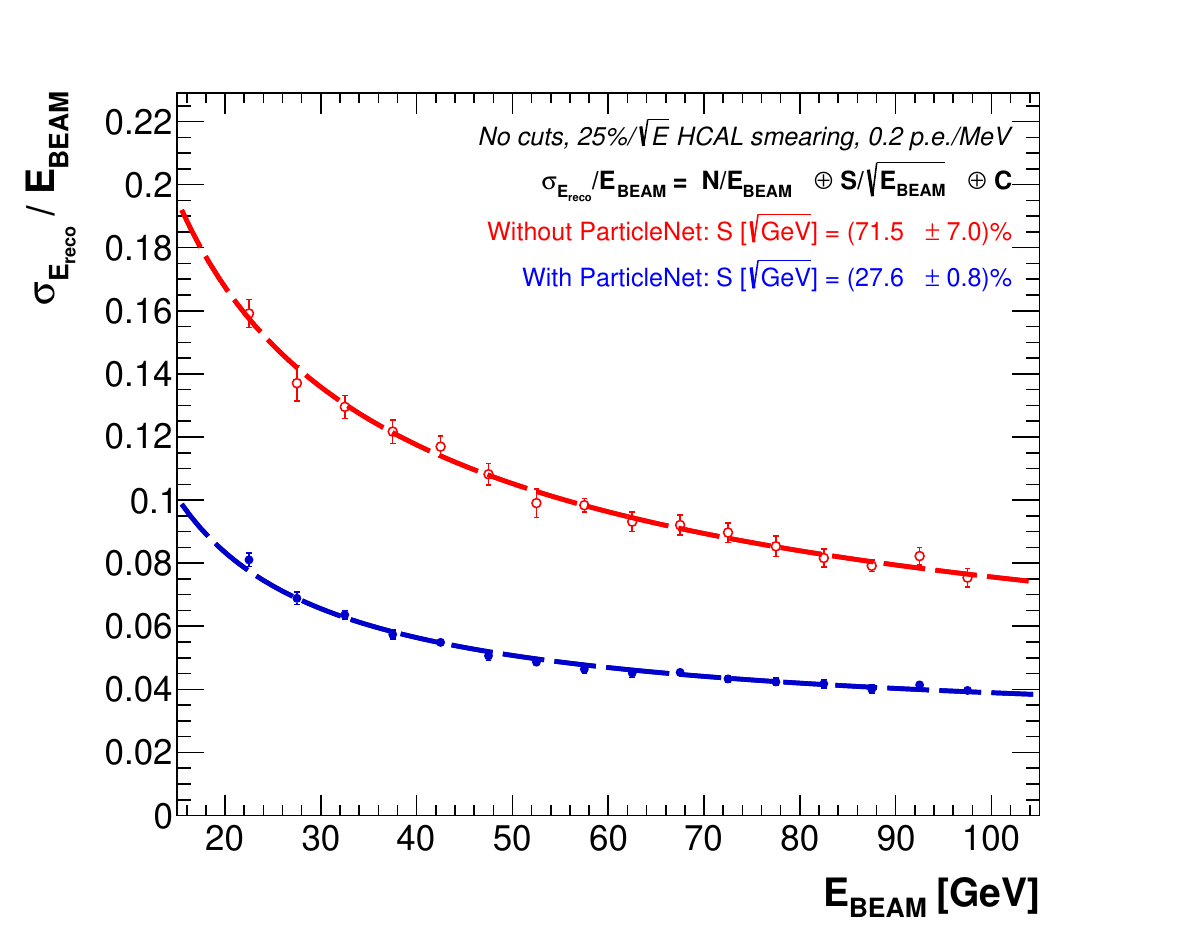}
\caption{Energy resolution for a simulated pion sample as a function of the beam energy, before and after GNN-based software compensation. The curves are fitted using a three-component model fit, described in the text.}
\label{fig:resognn}
\end{figure}

The energy resolution in different beam energy intervals for the test sample is shown in Figure~\ref{fig:resognn} as a function of beam energy.
The same three-component fit employed above has been used, yielding a stochastic term ($S\cdot\sqrt{E}$) of $(27.6 \pm 0.8)$\% after the correction. The constant term, $(2.49 \pm 0.08)\%$, is significantly larger than the one obtained for one-dimensional corrections. This could be related to the large degree of correlation between the stochastic and constant terms, but dedicated studies to address this type of effect are beyond the scope of this paper. For future $e^+e^-$ Higgs factories, the energy carried by individual hadrons within jets is typically much lower than 100~GeV, therefore this is not a limiting factor. Nonetheless, further studies with a dedicated high-energy simulated sample above 100~GeV may be required.
The noise term is $(1.02 \pm 0.05)$~GeV, which is compatible with the one found using the other approaches. 

\begin{figure}[!htbp]
\centering
\includegraphics[width=0.8\linewidth]{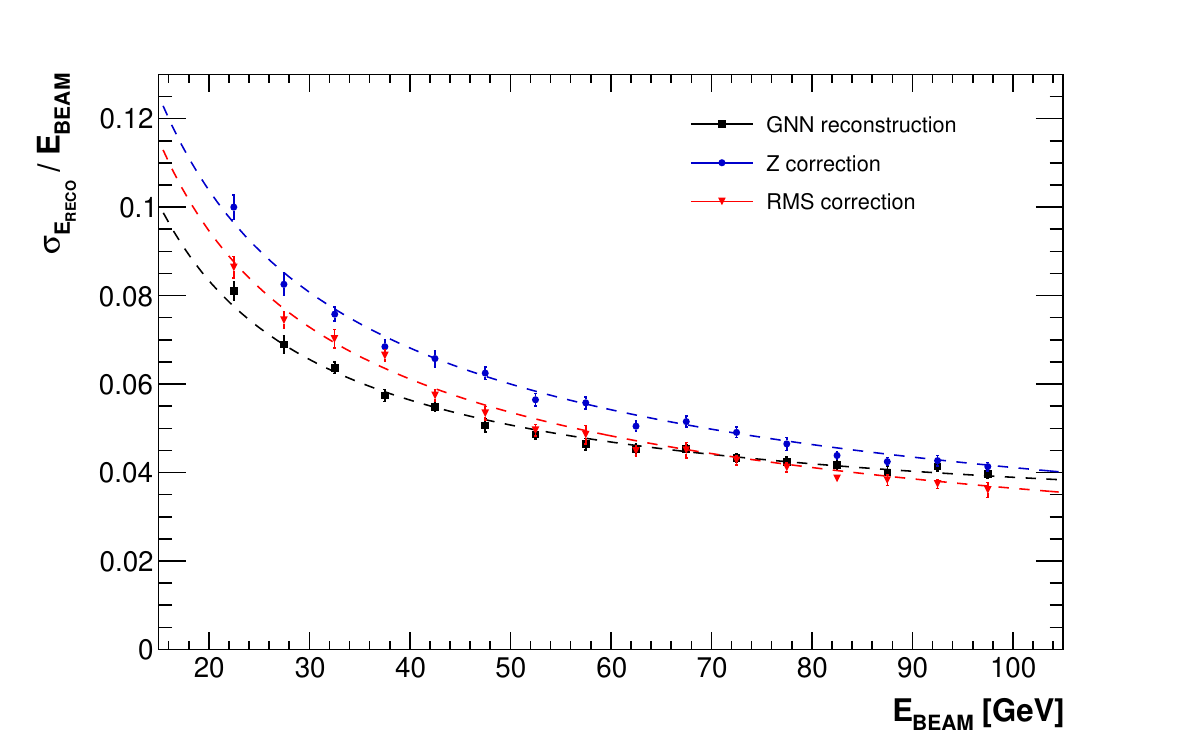}
\caption{Comparison of the energy resolutions with the RMS-based, the Z-based corrections, and the GNN reconstruction, for a simulated pion sample.}
\label{fig:compare}
\end{figure}

A comparison of the energy resolutions in the three approaches considered in the paper is shown in Figure~\ref{fig:compare}.
The fitted $N$, $S$ and $C$ coefficients of the energy resolution model are shown in Table~\ref{tab:tab-rms-gnn} for the two best-performing methods, namely RMS-based one-dimensional correction and GNN.

\begin{table}[!htbp]
\centering
\begin{tabular}{lcc}
\hline
 & With GNN & With RMS-based correction \\
\hline
$N$ [GeV] & $1.02 \pm 0.05$ & $1.01 \pm 0.02$ \\
$S$ [GeV$^{0.5}$] & $(27.6 \pm 0.8)\%$ & $(34.4 \pm 0.3)\%$ \\
$C$ & $(2.49 \pm 0.08)\%$ & $(0.52 \pm 0.02)\%$ \\
\hline
\end{tabular}
\caption{Values of the three components of the resolution model in the two cases with GNN reconstruction and with only RMS-based correction applied.}
\label{tab:tab-rms-gnn}
\end{table}

In Appendix~\ref{app:appendice-dr}, a comparison between the results obtained with this GNN and a dual-readout-based approach has been performed. Under the assumptions of this study, the dual-readout and GNN-based software compensation approaches applied in a CRILIN-like calorimeter achieve comparable resolutions, with relative differences typically below about 15\%.

\section{Extraction of the contribution from CRILIN to the combined resolution}

To quantify the contribution of CRILIN after GNN reconstruction to the combined CRILIN+HCAL energy resolution, the combined resolution was evaluated for each beam energy interval and the HCAL-only resolution was subtracted in quadrature.
This observable quantifies the extra resolution that would be added to a calorimetric system consisting of only a HCAL after including CRILIN upstream.

\begin{figure}
    \centering
    \includegraphics[width=0.85\linewidth]{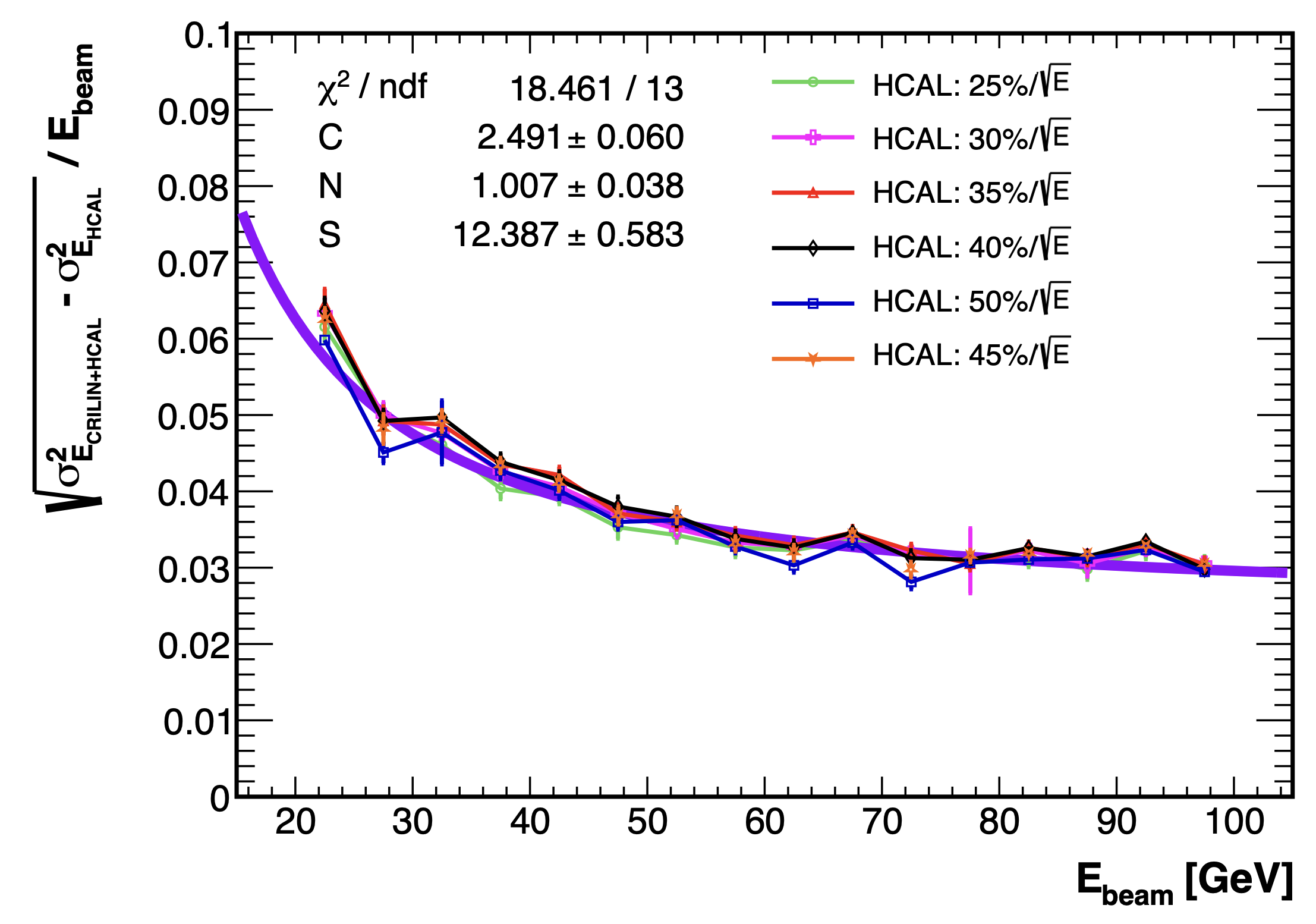}
    \caption{Contribution of CRILIN after GNN reconstruction to the combined CRILIN+HCAL energy resolution, obtained by subtraction in quadrature, as a function of the beam energy.}
    \label{fig:sm_compare}
\end{figure}

The impact of the choice of the downstream hadronic calorimeter resolution has been evaluated by varying the HCAL stochastic term ($S\cdot\sqrt{E}$) between 25\% and 50\%, as shown in Figure~\ref{fig:sm_compare}, and is found to have only a minor impact.
The contribution of CRILIN to the combined hadronic energy resolution is below 5\% for energies above 30~GeV, and it has been fitted with the same three-component function as already done above, yielding a noise term of $(1.007 \pm 0.038)$~GeV, a constant term of $(2.49 \pm 0.06)\%$, and a stochastic term of $12.39 \pm 0.58$ \%.


\section{Conclusions}

Software compensation techniques for the event-by-event correction of reconstructed hadron energies in the longitudinally segmented Cherenkov crystal CRILIN calorimeter have been investigated.
The study is based on Geant4 simulations of charged-pion hadronic showers and compares different types of one-dimensional and multivariate corrections.
The fine granularity of the detector provides access to shower-shape observables correlated with the fraction of total deposited energy that is reconstructed by the calorimeter, allowing the reconstructed energy to be corrected on an event-by-event basis.

Simple corrections based on global shower observables such as the transverse shower RMS or the longitudinal shower center-of-gravity already provide substantial improvements with respect to the uncorrected response. 

The physical origin of the RMS-based correction effectiveness was studied in detail, finding that the cluster RMS is correlated with the electromagnetic fraction and reproducing this behavior with a simplified phenomenological model employing two Gaussian shower sub-components, for the purely electromagnetic and hadronic parts.

A ParticleNet Graph Neural Network exploiting the full three-dimensional hit information provides a robust alternative to one-dimensional corrections, outperforming them for beam energies below approximately 70~GeV.

The contribution of the CRILIN calorimeter to the combined hadronic energy resolution has also been estimated by subtracting the assumed HCAL resolution in quadrature from the total combined resolution. 
The CRILIN contribution is found to be largely insensitive to the assumed HCAL resolution, modeled in the range between \(25\%/\sqrt{E_{HCAL}}\) and \(50\%/\sqrt{E_{HCAL}}\), and is described approximately by $
\frac{\sigma_E^{\mathrm{CRILIN}}}{E_{beam}}
=
1.01~\mathrm{GeV}/{E_{beam}}\,
\oplus\,
12.4\%/\sqrt{E_{beam}} \, \oplus \, 2.5\%$.
The corresponding contribution to the total calorimetric resolution is below approximately \(5\%\) for beam energies above \(30~\mathrm{GeV}\).
A dedicated study of the dependence on Geant4 physics lists is deferred to future work.

The ParticleNet-based reconstruction achieves a performance comparable to that obtained with dual-readout corrections under the assumptions considered in this work, by exploiting shower-shape information instead of separate scintillation and Cherenkov light measurements to recover information related to event-by-event fluctuations of the electromagnetic shower component.

These results demonstrate that a crystal calorimeter, despite its intrinsically non-compensating response, can recover a large fraction of the information lost by hadronic shower fluctuations through software compensation techniques. For this reason, with a proper HCAL and a CRILIN-like ECAL, excellent hadron energy resolution can be achieved, suggesting that the technique described could represent a viable alternative to dual-readout calorimetry in ECALs, in future experiments at $e^+e^-$ colliders. 


\clearpage
\begin{appendices}
\counterwithin{figure}{section}
    
\section{Studies on the Cherenkov response with a simulated muon sample}
\label{app:appendice-muoni}
\begin{figure}[!htbp]
\centering
\includegraphics[width=0.7\linewidth]{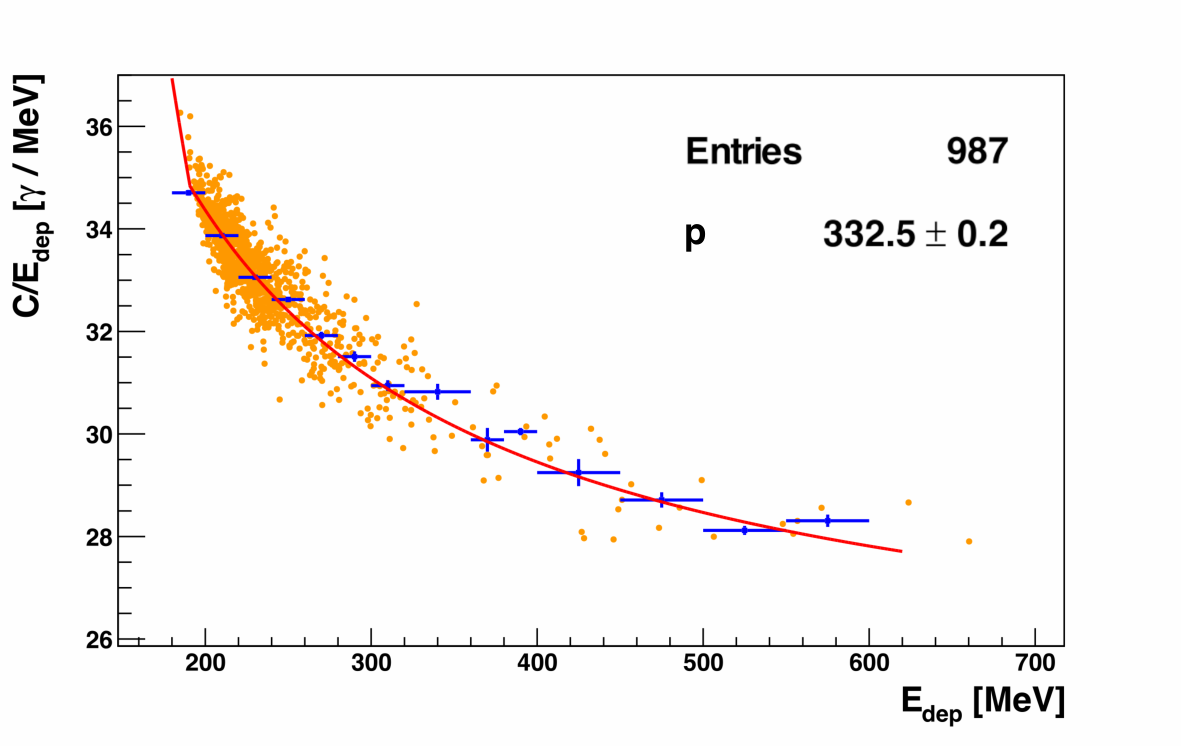}
\caption{Ratio between emitted Cherenkov photons and deposited energy $E_{dep}$, as a function of $E_{dep}$, for a 5~GeV muon sample, superimposed with a profile histogram. The fit function is described later in the text.}
\label{fig:mu-nonline}
\end{figure}

\begin{figure}[!htbp]
    \centering
    \includegraphics[width=0.43\linewidth]{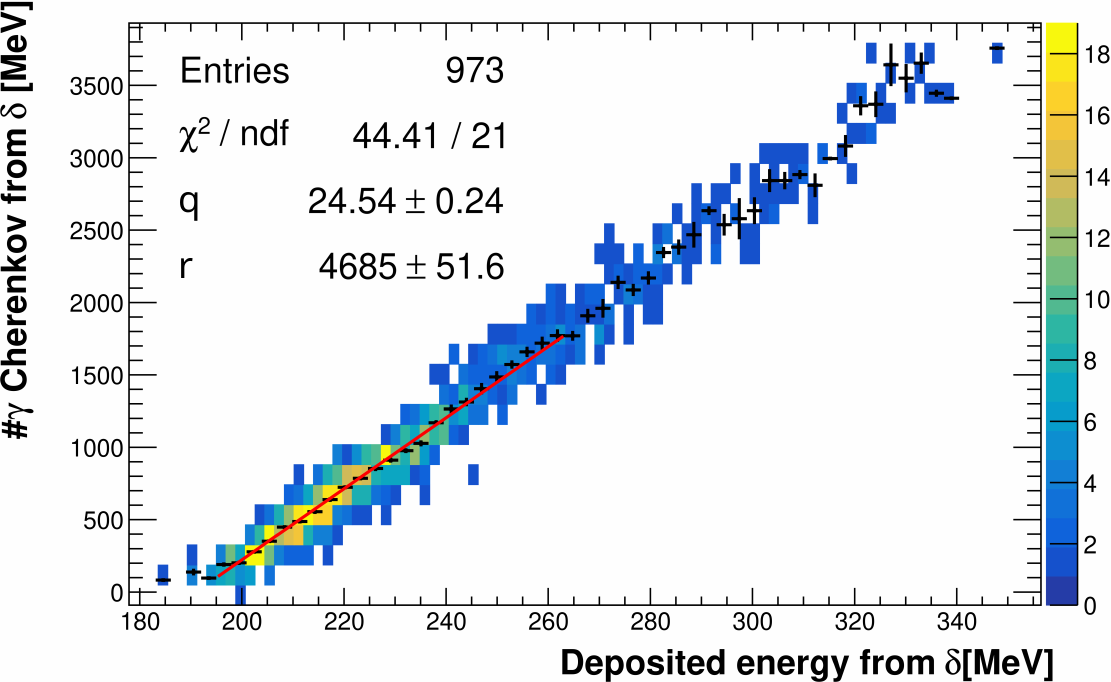}
    \hspace{2mm}
    \includegraphics[width=0.44\linewidth]{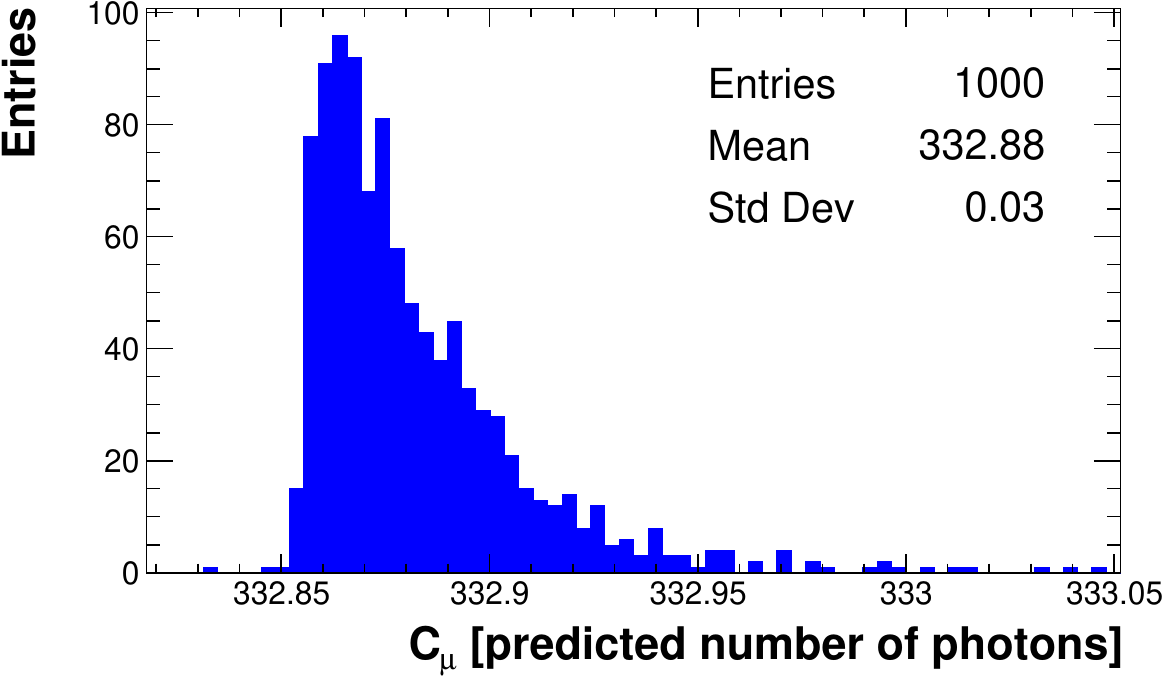}
    \caption{Left: distribution of \(C_{\delta}\) as a function of the energy deposited by delta rays, \(E_{\mathrm{dep}}\), summed over all delta rays in the event. Right: distribution of \(C_{\mu}\) normalized over the total crystal length. Both plots refer to a simulated 5~GeV muon sample.}
\label{fig:mu-delta}
\end{figure}

The origin of this non-linearity observed for muons lies in the Cherenkov response \(C\) which can be decomposed into two contributions. The first is due to the primary muon,
\[
C_{\mu} = p \,\Delta L ,
\]
which is simply proportional, through a constant coefficient \(p\), to the muon track length \(\Delta L\), as shown in Figure~\ref{fig:mu-delta}. The second contribution, denoted \(C_{\delta}\), arises from Cherenkov photons emitted by secondary delta rays.
Note that at the minimum ionizing particle (MIP) energy deposit, which is around 190~MeV for the simulated setup, the average $C/E_{dep}$ for muons is approximately 35~$\gamma$/MeV, corresponding to a relative difference of about 50\% in the muon and showering-electron conversion factor.
To study these two components separately, a dedicated simulation is performed, in which the contribution of the primary muon and that of secondary particles are identified using their Geant4 track IDs. With this procedure, \(C_{\delta}\) is found to scale linearly with \(E_{\mathrm{dep}}\), but only above a threshold \(E_{\mathrm{dep}}^{\,C_{\delta}>0}\). This threshold is approximately 190~MeV, consistent with the minimum-ionizing energy deposit by muons in the crystals, as shown in Figure~\ref{fig:mu-delta}, where a linear fit with slope $q$ and offset $(- r)$ is performed. By contrast, \(C_{\mu}\) exhibits a very small relative spread, of the order of \(10^{-4}\), as also shown in Figure~\ref{fig:mu-delta}, consistent with the hypothesis of depending only on the practically constant track length.

The ratio between the Cherenkov response, \(C\), and the deposited energy, \(E_{\mathrm{dep}}\), can then be written for muons as
\begin{align}
\frac{C_\mu + C_\delta}{E_{\mathrm{dep}}}
&=
\frac{p \,\Delta L}{E_{\mathrm{dep}}}
+
\frac{\max\!\left(0, q E_{\mathrm{dep}}- r\right)}{E_{\mathrm{dep}}}
\nonumber \\
&=
\frac{p \,\Delta L}{E_{\mathrm{dep}}}
+
\max\!\left(0, q-\frac{ r}{E_{\mathrm{dep}}}\right),
\label{eq:muon_cherenkov_ratio}
\end{align}
where \(p\), \(q\), and \(r\) are positive coefficients satisfying
\[
 q E_{\mathrm{dep}}^{\,C_\delta > 0} -  r = 0.
\]

For \(E_{\mathrm{dep}} > E_{\mathrm{dep}}^{\,C_\delta > 0}\), the function becomes a hyperbola shifted by the constant term \( q\), and therefore \(C/E_{\mathrm{dep}}\) asymptotically approaches \( q\) at large \(E_{\mathrm{dep}}\).
The coefficient \(p\) is determined from a fit to the simulated muon sample using Equation~\ref{eq:muon_cherenkov_ratio}, as shown in Appendix~\ref{app:appendice-muoni}, Figure~\ref{fig:mu-nonline}, after fixing $q$ and $r$ to the values fitted in Figure~\ref{fig:mu-delta}. 
Note that the fitted value of $p$ is compatible with the average of the distribution of the number of emitted Cherenkov photons from the primary muon per cm in Figure~\ref{fig:mu-delta}, and also with the prediction of the Frank--Tamm formula in the wavelength range considered in the simulation.
Note also that the fitted value of the $q$ coefficient is compatible with the conversion factor obtained from electrons, for which most of the deposited energy is carried by relativistic electrons with energies around the critical energy.

To take this non-linear effect into account, throughout the paper, pions that do not start any hadronic shower in the CRILIN module (dubbed MIP events) are treated as muons, and the proper $C/E_{dep}$ function is employed to convert the Cherenkov response to an energy value, while for showering pions the electron-based conversion factor is used. This choice requires an efficient cut-based selection to separate MIP events, and the one employed in the following selects events with less than 5~GeV true deposited energy in CRILIN, with hits in a single column of crystals, with the least energetic hit greater than 30\% of the hit energy averaged in the 5 layers ($E_{avg}$), and with the most energetic hit not larger than 180\% of $E_{avg}$.

\section{Study of the correlation between the cluster width and \texorpdfstring{$f_{EM}$}{fEM}}
\label{app:appendice-rms}

\begin{figure}
    \centering
    \includegraphics[width=0.7\linewidth]{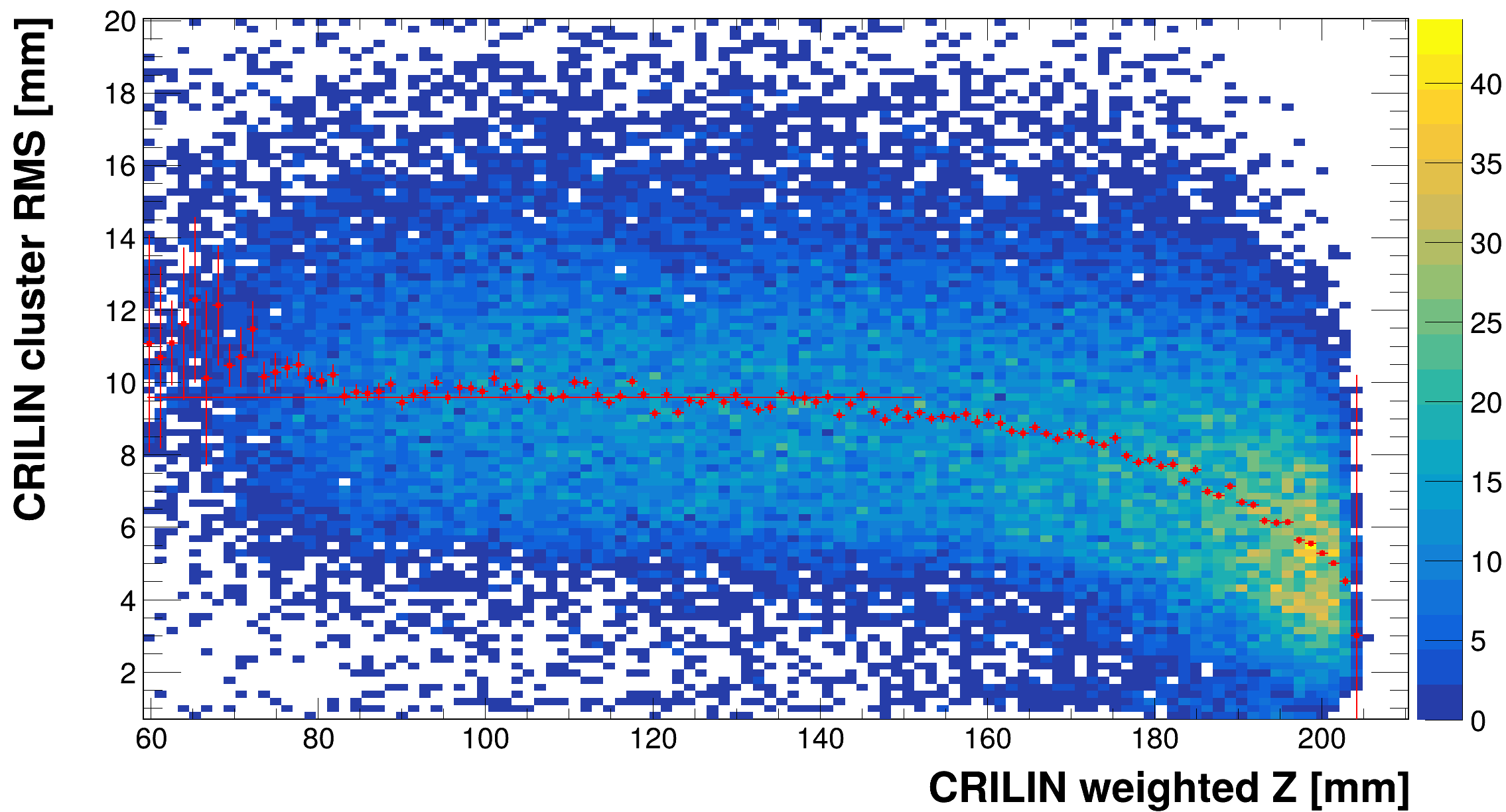}
    \caption{CRILIN cluster RMS for pion showers, as a function of the energy-weighted Z coordinate, evaluated using the energy barycenter on the longitudinal axis as detailed in Section~\ref{z}.}
    \label{fig:rms-z}
\end{figure}

It is interesting to motivate the linear correlation between the cluster RMS and $E^{reco}_{CRILIN}/E^{true}_{CRILIN}$ found in Section~\ref{rms} from a physical standpoint.
The cluster RMS of pion-induced showers depends on the interaction depth, as shown in Figure~\ref{fig:rms-z}, but its average value of about 10~mm is approximately constant up to $Z=150$~mm, where edge effects are present. For this reason, in order to reduce Z-dependent effects associated with CRILIN's edges, only events with $Z < 150$~mm are taken into account in this section. The simulated pion beam energy employed for this study is uniformly distributed between 15 and 35~GeV, restricting the study to a relevant but relatively small energy range.

A pion-initiated shower can be decomposed into a purely electromagnetic component, originating from $\pi^0/\eta\to \gamma\gamma$ decays, and a remaining hadronic component; the total energy readout with Cherenkov light in CRILIN ($C$) can be written as~\cite{wigmans-dual-readout}:
\begin{equation}
    C = E_{EM}^{C} + E_{had}^{C} = E [f_{EM} + (1 - f_{EM})\cdot (h/e)_{C}] ,
    \label{eq:c_fem}
\end{equation}
where $E$ is the true energy deposited in CRILIN, $E_{EM(had)}^{C}$ is the total energy of the electromagnetic (hadronic) component readout via Cherenkov light (C), and $f_{EM}$ is the electromagnetic fraction of the shower, which fluctuates on an event-by-event basis.
The parameter $(h/e)_{C}$ is the ratio between the detector responses of the hadronic and electromagnetic shower components.

Consequently, the RMS of the shower can be decomposed into these two components, by approximating the shower as a Gaussian mixture of the two components, weighted in energy:

\begin{equation}
    RMS = \sqrt{\frac{E_{EM}^{C} RMS^2_{e} + E_{had}^{C} RMS^2_{had}}{E_{EM}^{C} + E_{had}^{C}}} = \sqrt{\frac{f_{EM} RMS^2_{e} + (1 - f_{EM})\cdot (h/e)_{C} \cdot RMS^2_{had}}{f_{EM} + (1 - f_{EM})\cdot (h/e)_{C}}}.
    \label{eq:rms_c}
\end{equation}

The parameter $RMS_{had}$ is operationally defined as the RMS of a CRILIN cluster with no $\pi^0$ or $\eta$ production ($f_{EM}=0$).
The expression in Equation~\eqref{eq:rms_c} should be regarded as a simplified phenomenological model rather than a rigorous decomposition of hadronic showers.

\begin{figure}[!htbp]
    \centering
    \includegraphics[width=0.7\linewidth]{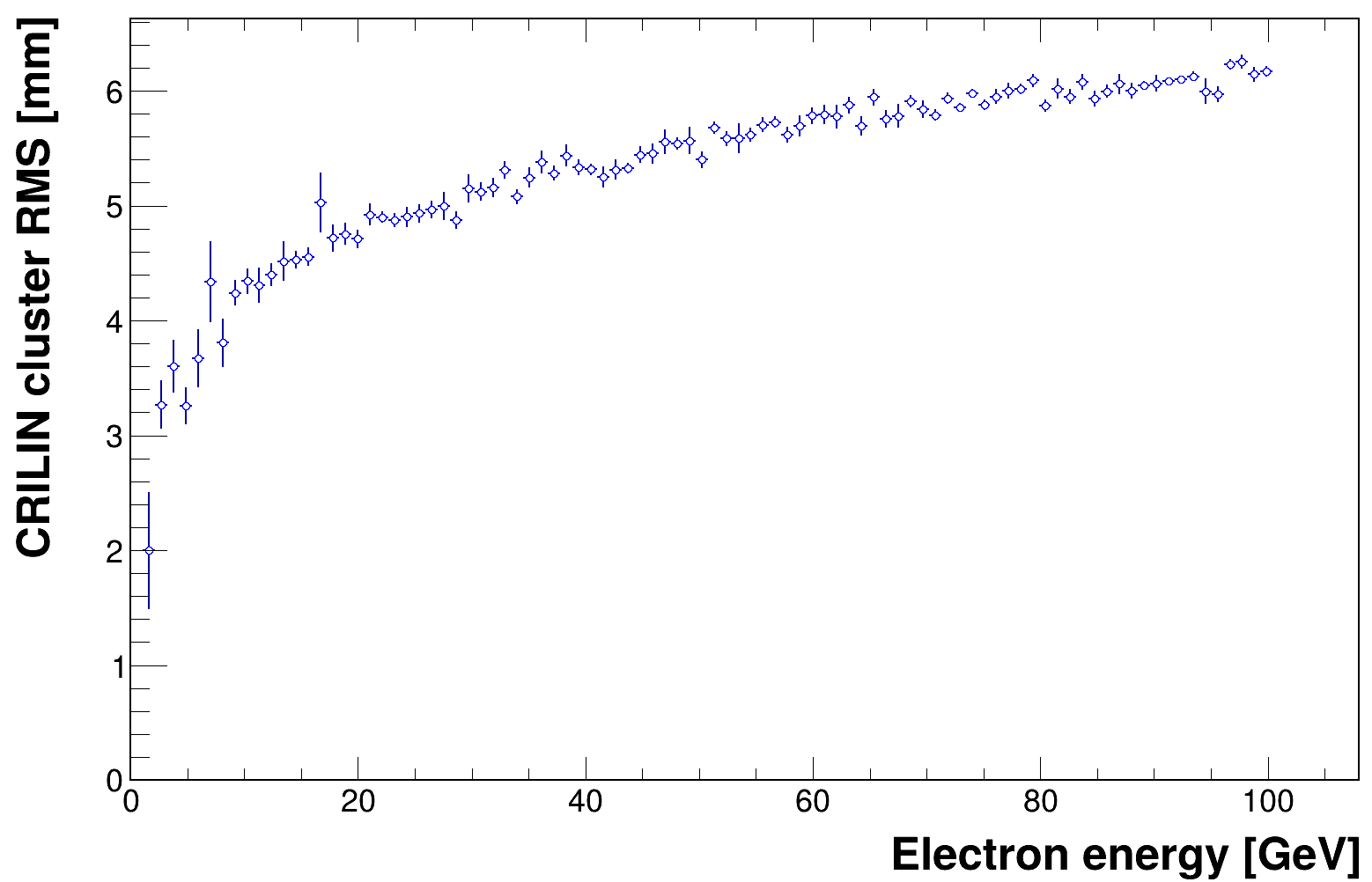}
    \caption{CRILIN cluster RMS for electron-induced showers as a function of beam energy.}
    \label{fig:ele-rms}
\end{figure}

Electron-initiated clusters have a smaller RMS in CRILIN compared to to pion-initiated showers, whose average RMS is approximately 10~mm.
As shown in Figure~\ref{fig:ele-rms}, the average value of $RMS_{e}$ is about 4.5~mm at 15~GeV, 5~mm at 30~GeV and 6~mm at 100~GeV beam energy.

\begin{figure}[!htbp]
    \centering
    \includegraphics[width=0.7\linewidth]{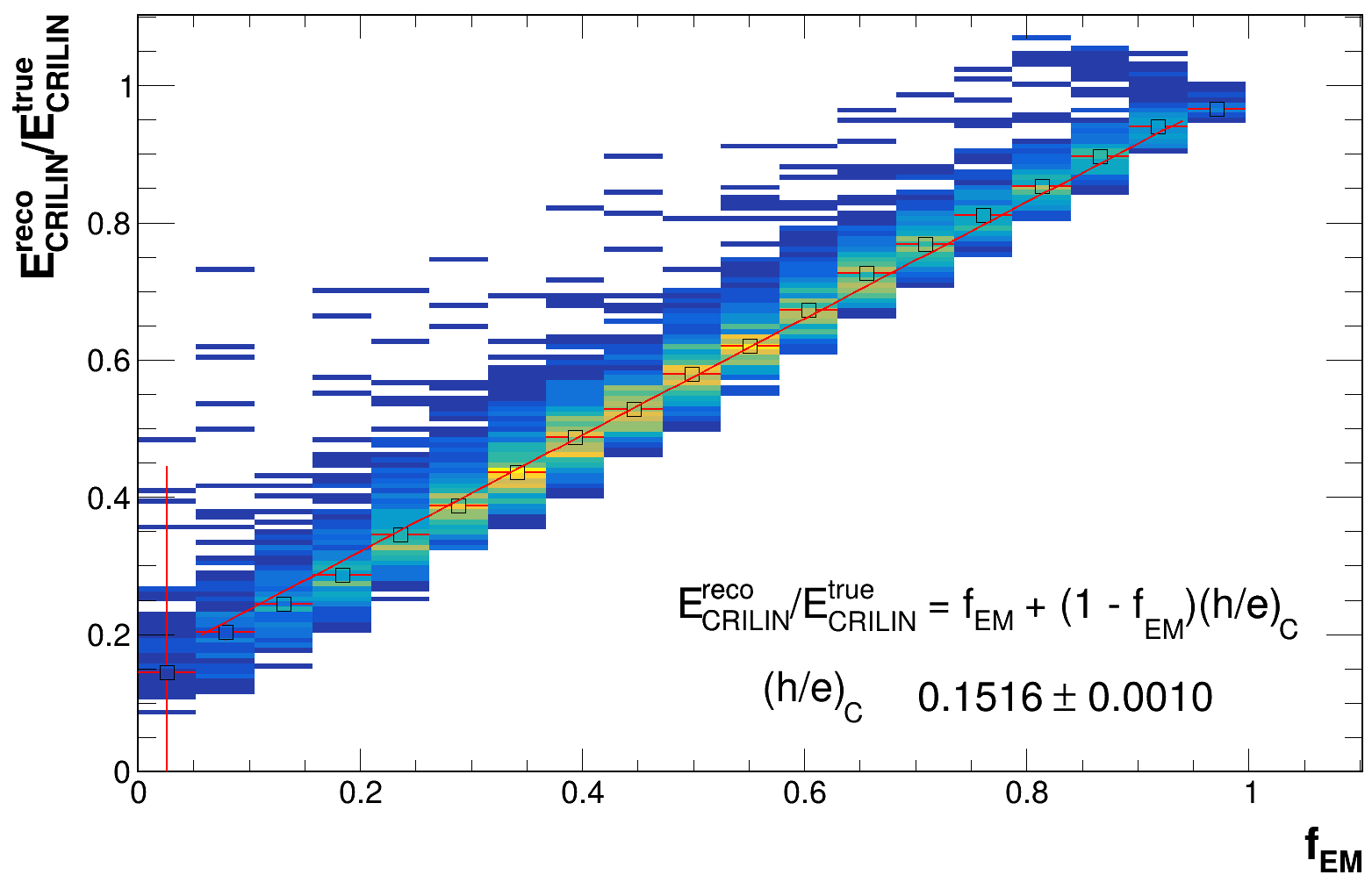}
    \caption{Ratio $C/E = E_{CRILIN}^{reco}/E_{CRILIN}^{true}$, as a function of $f_{EM}$, fitted using Equation~\eqref{eq:c_fem}.}
    \label{fig:h_over_e_c}
\end{figure}

The ratio $(h/e)_{C}$ can be determined with a dedicated simulation, where $f_{EM}$ is recorded for each event. For this reason, the electromagnetic fraction $f_{EM}$ is estimated by summing all energy deposits coming from electrons, photons and positrons, which, to a first approximation, are only due to the electromagnetic showers following $\pi^0$ or $\eta$ decays.
As shown in Figure~\ref{fig:h_over_e_c}, the fitted value from simulation is $(h/e)_{C} \approx 0.15$, indicating that only 15\% of the energy released hadronically, i.e. that is not released via electromagnetic showers, is detected through Cherenkov light.

\begin{figure}[!htbp]
    \centering
    \includegraphics[width=0.7\linewidth]{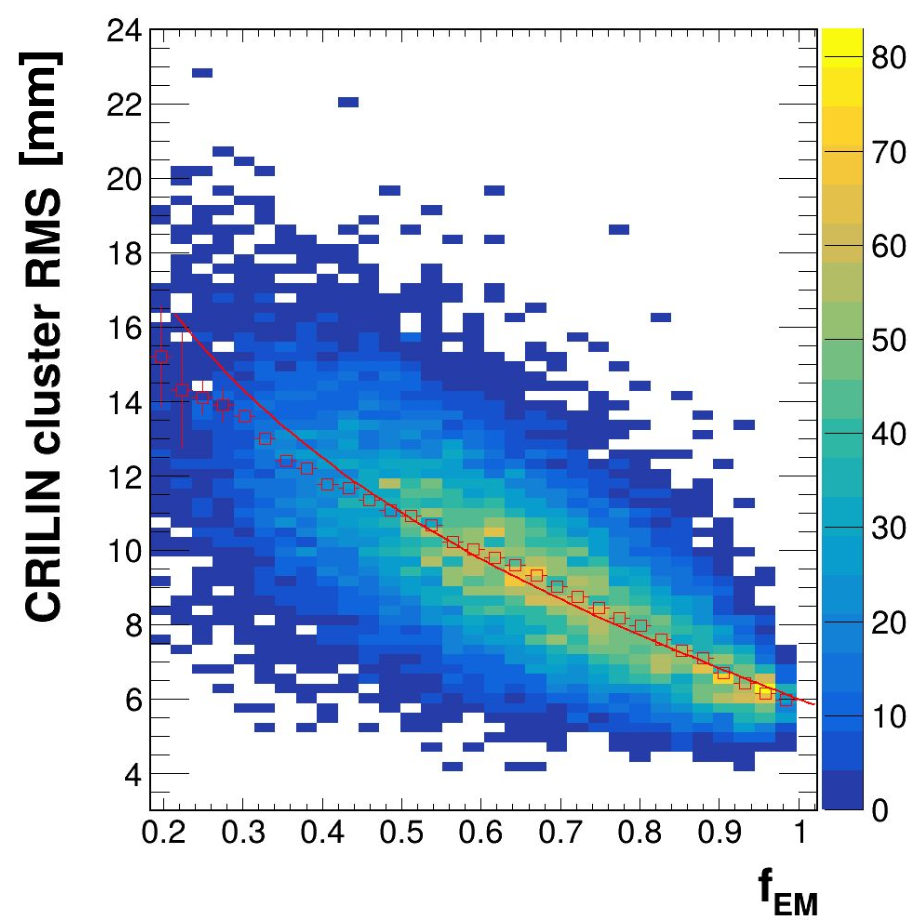}
    \caption{Distribution of cluster RMS as a function of $f_{EM}$, with superimposed profile histogram and fit function from Equation~\eqref{eq:rms_c}.}
    \label{fig:rms_fem_fit}
\end{figure}

Knowing the value of $(h/e)_{C}$, it is possible to verify the validity of Equation\eqref{eq:rms_c} and fit $RMS_{e}$ and $RMS_{had}$.
As shown in Figure~\ref{fig:rms_fem_fit}, the model is compatible with the simulated data to a level below 10\%, which is sufficient for the scope of this paper. 

The fitted values of $RMS_{e}$ and $RMS_{had}$ are about 5.5~mm, and 26~mm, respectively. The fit function has only a small curvature over the considered range, while the simulated data show an even weaker curvature.
This justifies the use of the linear corrections introduced in Section~\ref{rms} as an approximation, due to the linear dependence of $C/E$ on $f_{EM}$ described by Equation~\eqref{eq:c_fem}.

\section{Comparison with the dual-readout approach}
\label{app:appendice-dr}

To compare software compensation in CRILIN with the well-established dual-readout compensation approach, the results obtained with the GNN are compared with those reported in~\cite{lucchini}. These were obtained from simulations of neutral hadrons in a dual-readout PbWO$_4$ crystal ECAL followed by a dual-readout HCAL, for which a stochastic term of $25\%/\sqrt{E}$ was assumed.

Under the different assumptions considered in the two studies, the performance reported in~\cite{lucchini} is comparable to that obtained with the GNN in the present work, with relative differences below approximately 15\%.

The two studies differ in several aspects, such as detector geometry, crystal type, amount of passive material, and treatment of the HCAL; the present work also uses a more conservative definition of energy resolution, as explained in Section~\ref{rms}.
For this reason, a dedicated study has been performed using the simulation of the CRILIN calorimeter, by recording the amount of scintillation light released in each crystal separately from the information on the Cherenkov light. The light yield considered is 2 p.e./MeV, which is consistent with the one of PbWO$_{4}$. 
Since CRILIN is a pure Cherenkov-based calorimeter, and PbF$_{2}$ is a non-scintillating crystal, the scintillation light added is a non-physical feature, which serves only to illustrate the performance achievable with a dual-readout implementation in a CRILIN-like calorimeter, and to compare it with the GNN-based results. In principle, the simulation should have been performed with PbWO$_{4}$, but this would have required a new training of the GNN, which is outside the scope of this study. Note also that PbWO$_{4}$ and PbF$_{2}$ have similar nuclear properties, with radiation (nuclear interaction) lengths of, respectively, 0.89 (20.27)~cm, and 0.94 (22.10)~cm.

\begin{figure}[!htbp]
    \centering
    \includegraphics[width=0.8\linewidth]{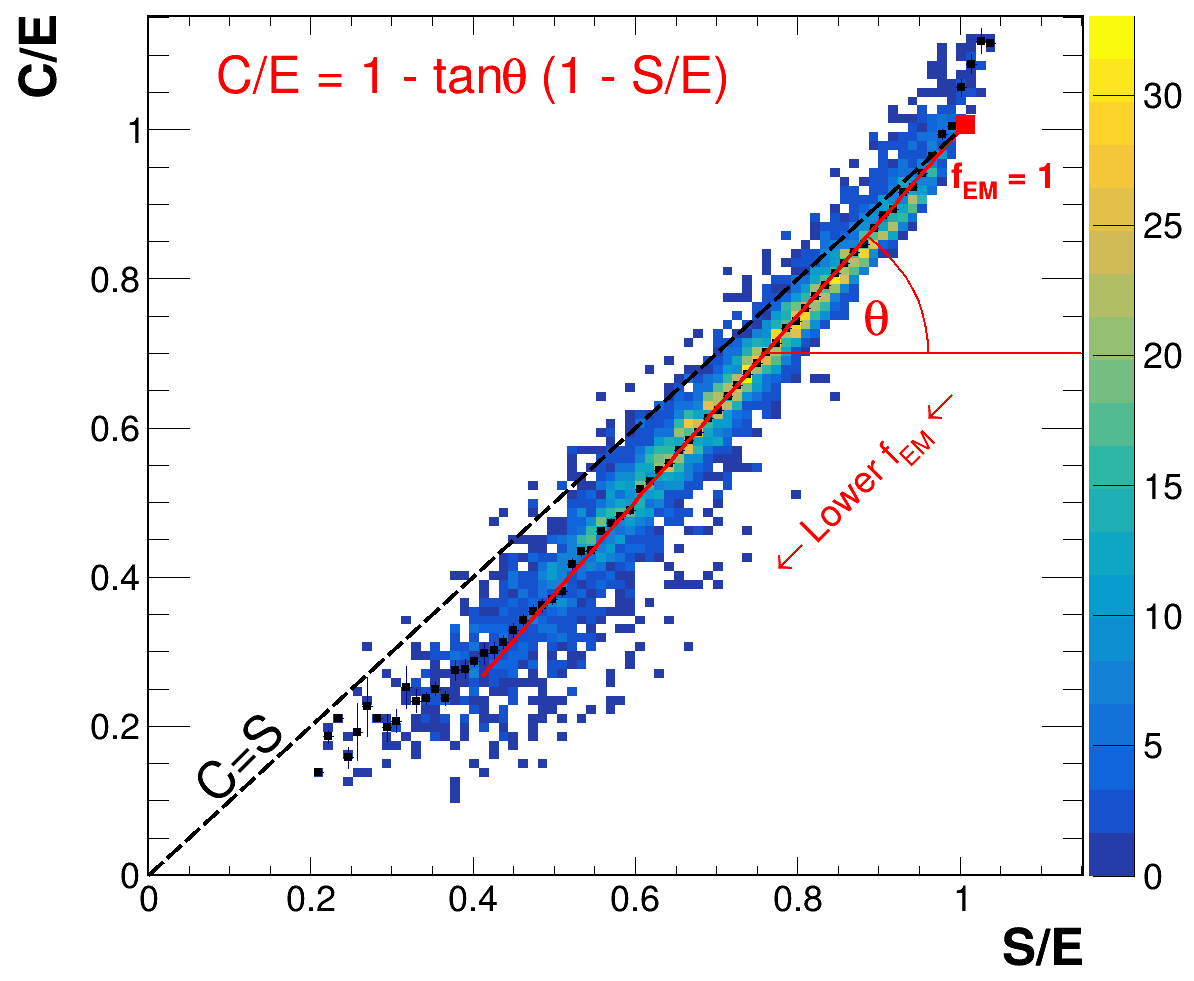}
    \caption{Joint distribution of $(C/E)$ vs. $(S/E)$, superimposed with the $C=S$ line and a linear fit to data. The point at $f_{EM}=1$ is shown, together with the $\theta$ angle.}
    \label{fig:dr}
\end{figure}

The dual-readout implementation follows closely the technique shown in~\cite{wigmans-dual-readout}, where for any true energy $E$, $C$ (Cherenkov light total equivalent energy) and $S$ (scintillation light total equivalent energy), both calibrated with electrons, are taken into account, and the joint distribution of $C/E$ and $S/E$ is employed to define an angle $\theta$ between the line where the data points accumulate and the $(S/E)$ axis, as shown in Figure~\ref{fig:dr}.

The angle $\theta$ is used to define $\chi =\cot(\theta) \approx 0.803$, which is the only parameter needed to determine the true energy~\cite{wigmans-dual-readout} using the equation:
\begin{equation}
    E = \frac{S - \chi C}{1 - \chi}.
\end{equation}

\begin{figure}[!htbp]
    \centering
    \includegraphics[width=0.75\linewidth]{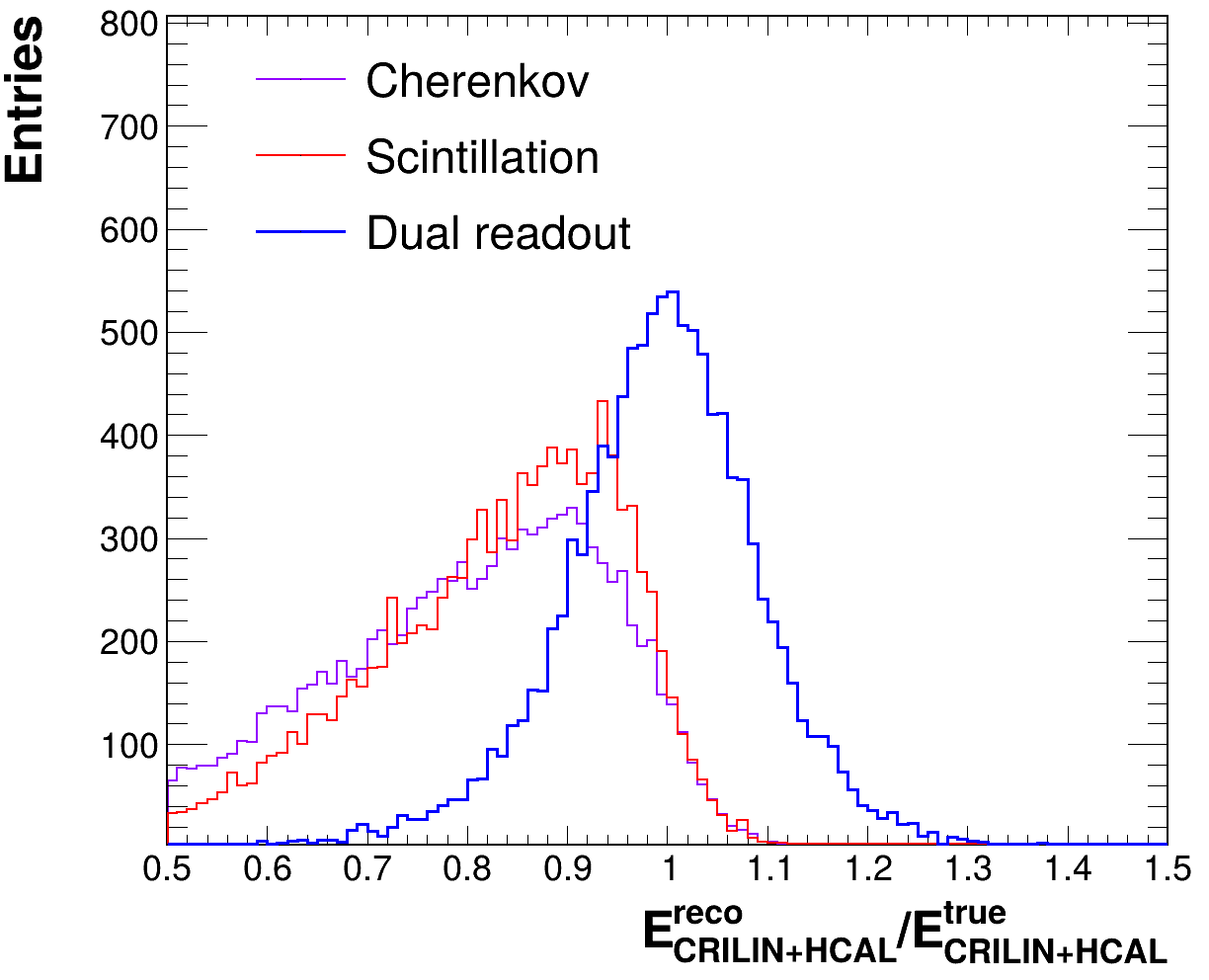}
    \caption{Distribution of the ratio between total reconstructed energy and total true energy in CRILIN + HCAL, considering only events where the shower starts in the ECAL, taking into account scintillation light (red), Cherenkov light (violet) and using a dual-readout correction (blue). The HCAL energy is evaluated with a parametric $25\%/\sqrt{E}$ smearing applied on the energy readout with a virtual detector downstream of CRILIN.}
    \label{fig:dr_corr}
\end{figure}

The effect of the dual-readout energy correction is shown in Figure~\ref{fig:dr_corr}, for an HCAL energy smearing of $25\%/\sqrt{E}$. 

\begin{figure}[!htbp]
    \centering
    \includegraphics[width=0.8\linewidth]{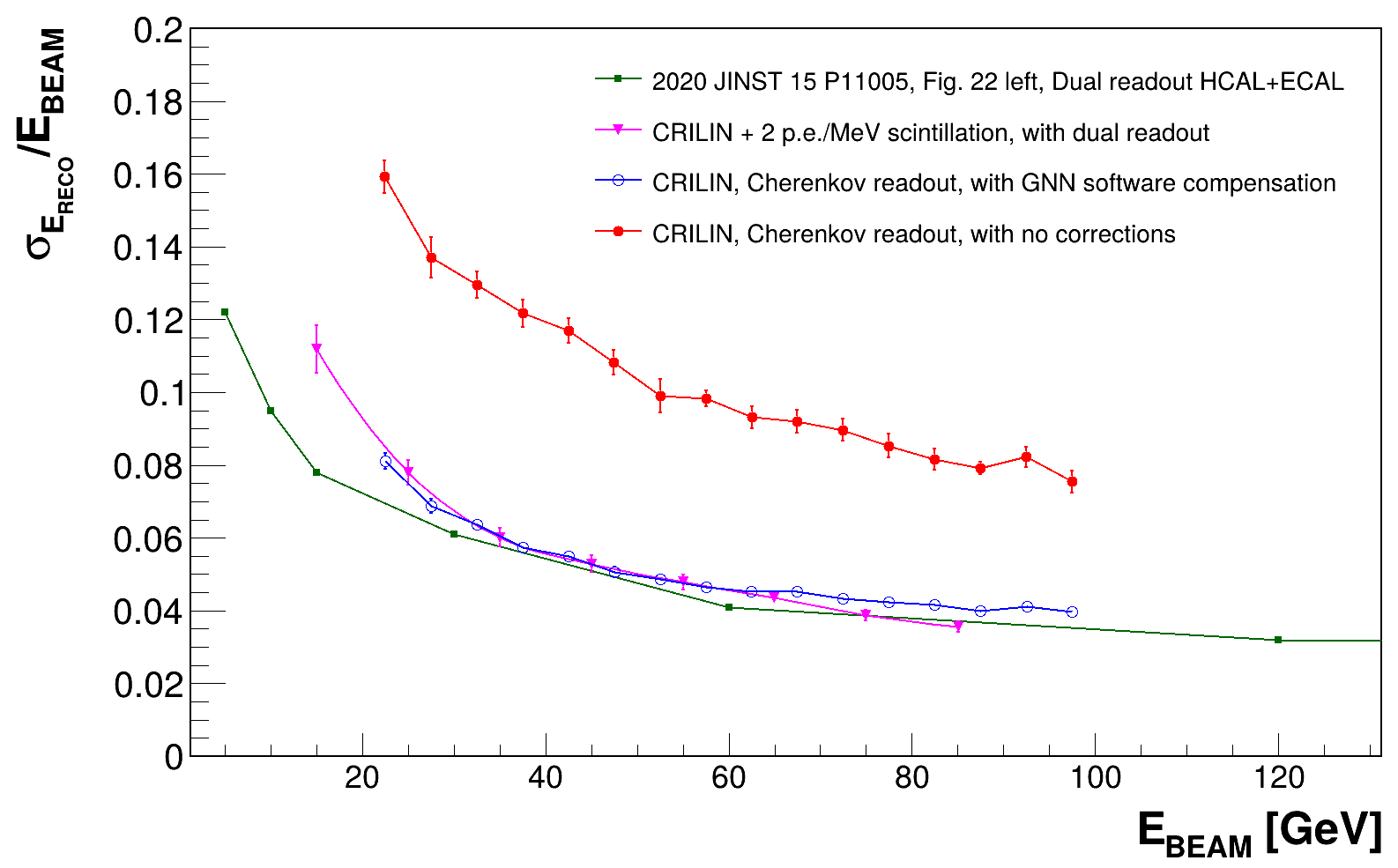}
    \caption{Energy resolution of the ECAL+HCAL system as a function of hadron beam energy. The curves shown correspond, respectively, to CRILIN with no corrections (red), CRILIN with dual-readout (pink), CRILIN with GNN corrections (blue), and a PbWO$_{4}$ crystal calorimeter with dual-readout~\cite{lucchini} (green). For the latter curve, the HCAL energy is evaluated with a full simulation, while for CRILIN a parametric $25\%/\sqrt{E}$ smearing is applied on the energy readout with a virtual detector.}
    \label{fig:dr_comp}
\end{figure}

The resulting energy resolution is shown in Figure~\ref{fig:dr_comp}, where the performances are found to be similar to the GNN-based corrections below 60~GeV.

\section{Study of the RMS-based corrections using scintillation light}
\label{app:appendice-scint}

\begin{figure}[!htbp]
    \centering
    \includegraphics[width=0.7\linewidth]{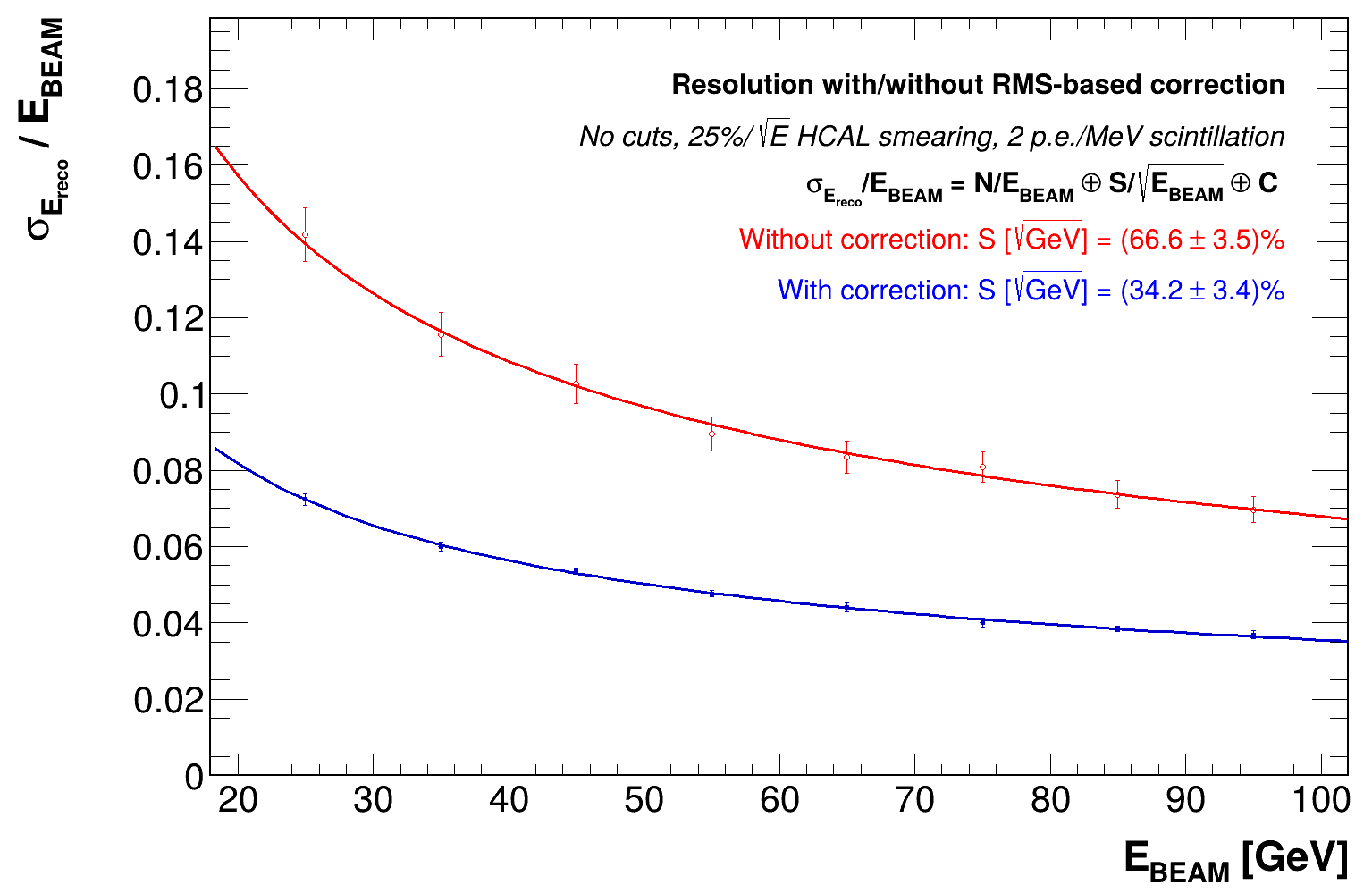}
    \caption{Energy resolution for a combined CRILIN and HCAL system, as a function of the beam energy of the simulated pion sample, with and without RMS-based correction applied. The CRILIN energy is evaluated using the non-physical scintillation output of a dedicated simulation, as detailed in the text.}
    \label{fig:rms_s}
\end{figure}

Using the simulation detailed in Appendix~\ref{app:appendice-dr} including non-physical scintillation light in PbF$_{2}$, the RMS-based corrections have been evaluated using the scintillation output instead of the Cherenkov light, to study whether the higher light yield leads to any improvement in performance.
The results are compatible within uncertainties with those obtained in Section~\ref{rms} with pure Cherenkov response, and are shown in Figure~\ref{fig:rms_s}, suggesting that, in the context of offline compensation, the performance gain obtained with a scintillating crystal calorimeter relative to a Cherenkov-based calorimeter is minimal.

\section{Study on the combination of RMS and Z-based corrections}
\label{app:appendice-z}

\begin{figure}[!htbp]
    \centering
    \includegraphics[width=0.75\linewidth]{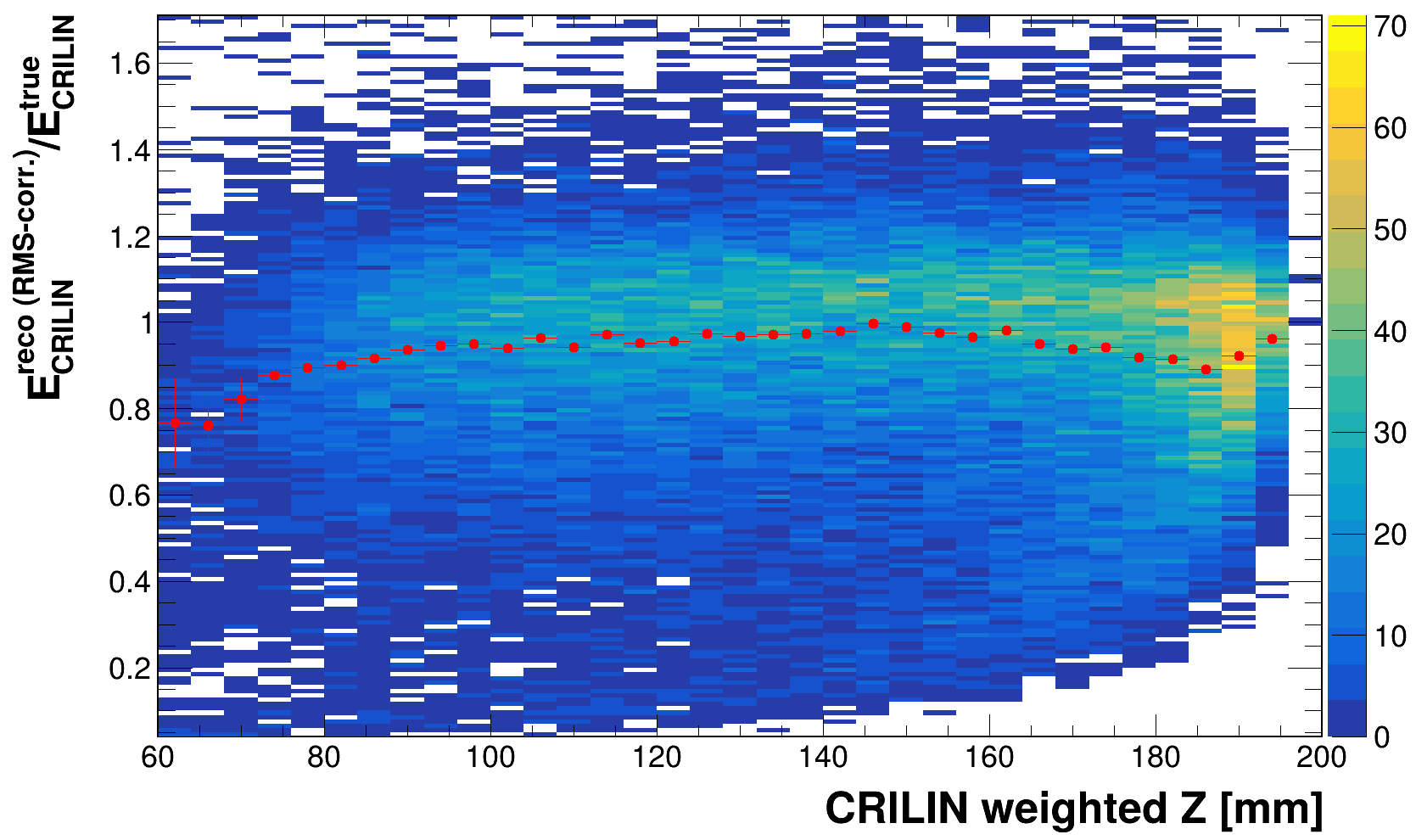}
    \caption{Joint distribution of the ratio between the energy reconstructed in CRILIN after RMS-based corrections and the true energy deposited in CRILIN, and the energy-weighted Z coordinate.}
    \label{fig:e_vs_z_z_dopo_rms}
\end{figure}

It is worth considering the possibility of applying RMS and Z-based corrections sequentially. This can be preferred over more complex multivariate corrections, i.e. the GNN-based ones, if the correlation between RMS and Z is mild, which is however true only for a fraction of the events, as shown in Figure~\ref{fig:rms-z}.
Nonetheless, the first step of this feasibility study has been carried out, by plotting the joint distribution of the ratio between the energy reconstructed in CRILIN after RMS-based corrections and the true energy deposited in CRILIN, and the energy-weighted Z coordinate. As shown in Figure~\ref{fig:e_vs_z_z_dopo_rms}, the energy ratio does not show a clear linear trend.  

\begin{figure}[!htbp]
    \centering
    \includegraphics[width=0.49\linewidth]{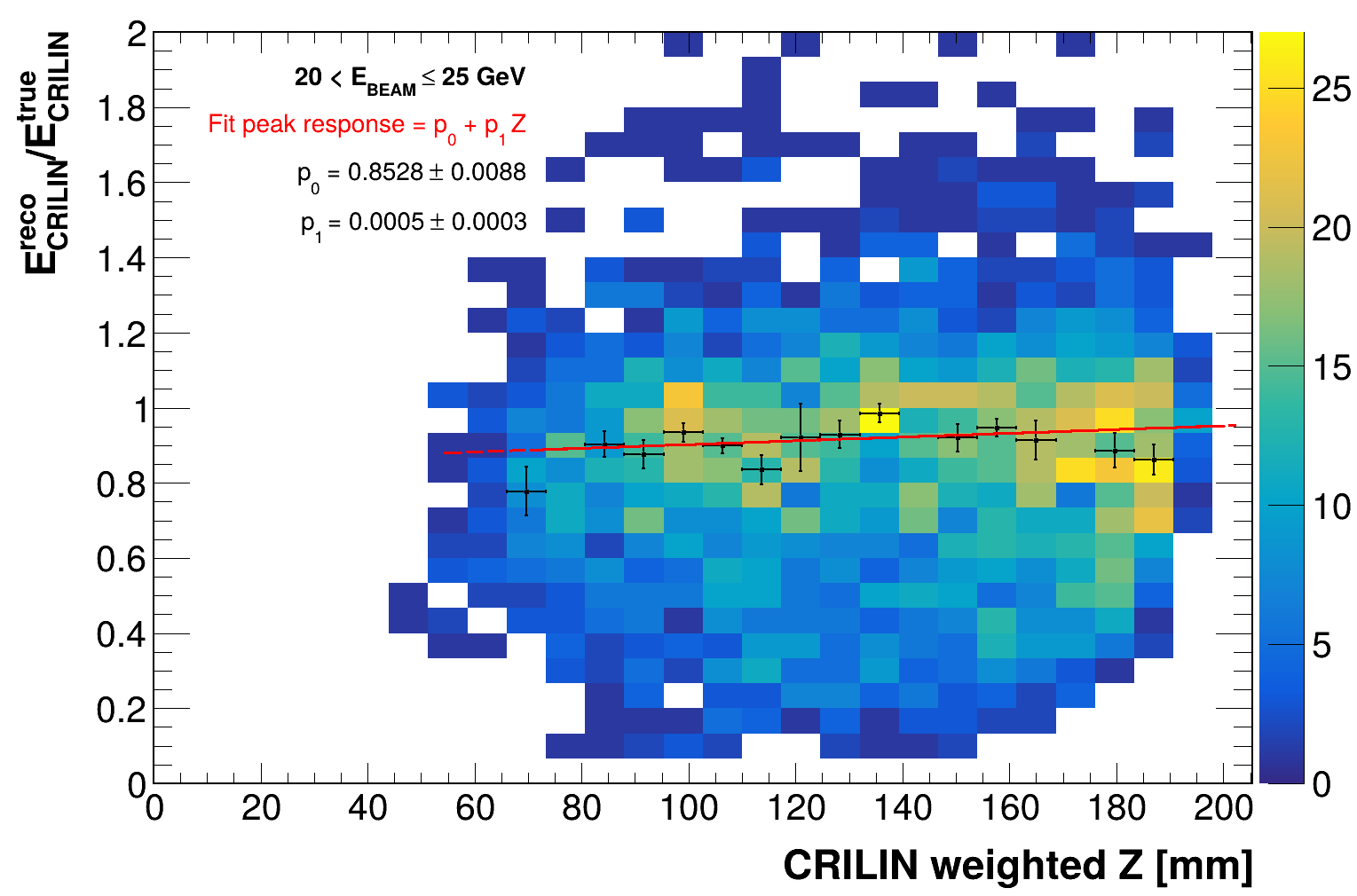}
        \includegraphics[width=0.49\linewidth]{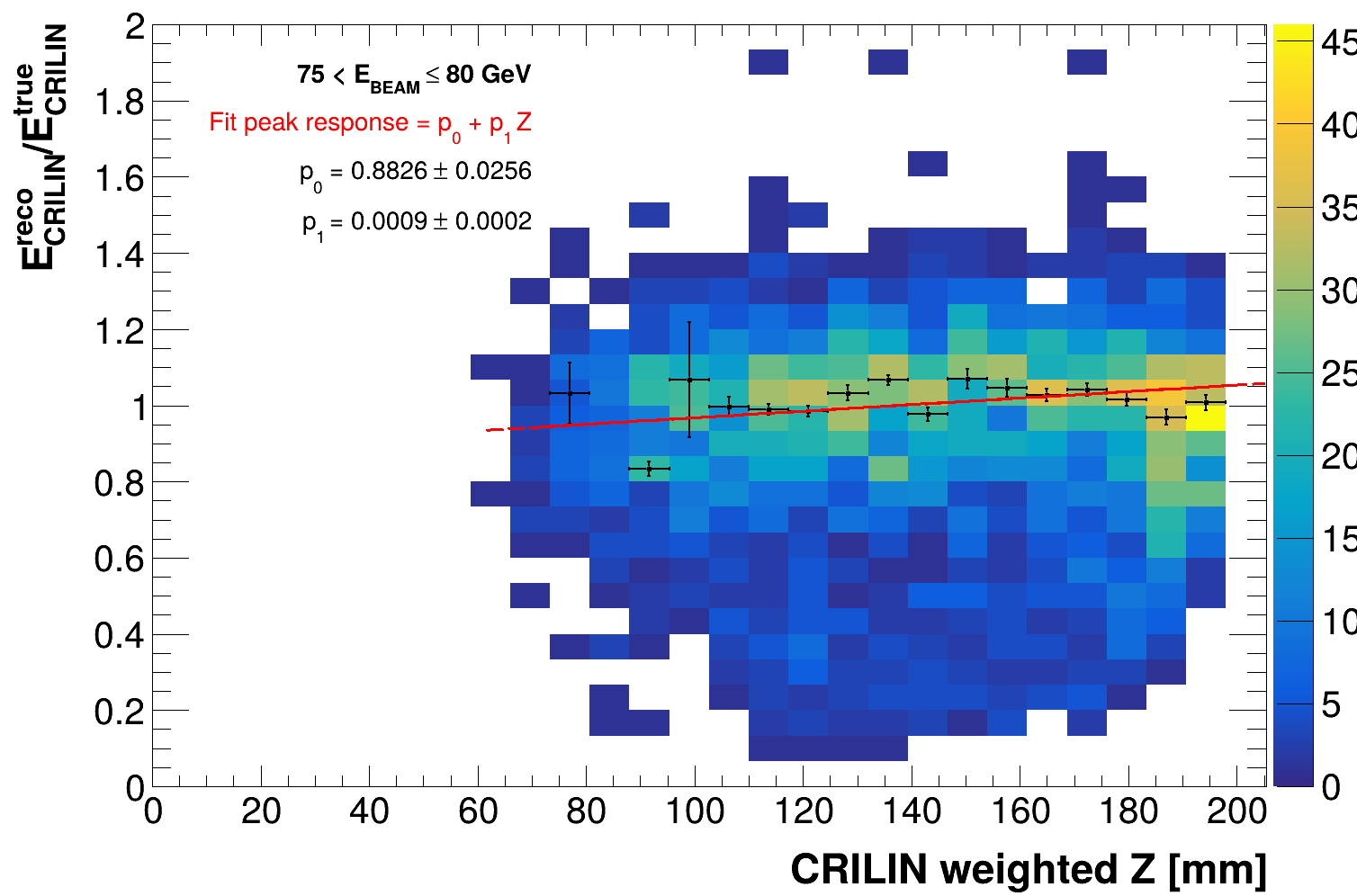}
    \caption{Joint distribution of the ratio between the energy reconstructed in CRILIN after RMS-based corrections and the true energy deposited in CRILIN, and the energy-weighted Z coordinate, in two beam energy slices, i.e. from 20 to 25~GeV (left), and from 75 to 80~GeV (right).}
    \label{fig:rms_plus_z_twoslices}
\end{figure}

However, the energy ratios could be fitted with separate linear functions in different beam energy ranges, using the same procedure as in Section~\ref{z}. Figure~\ref{fig:rms_plus_z_twoslices} shows the resulting fits for the same beam energy slices considered in Figure~\ref{fig:z-corr}.

The fitted linear functions have significantly smaller slopes with respect to the ones considered in Section~\ref{z}, because the largest part of the correction has already been applied with the RMS-based procedure.

\begin{figure}[!htbp]
    \centering
    \includegraphics[width=0.75\linewidth]{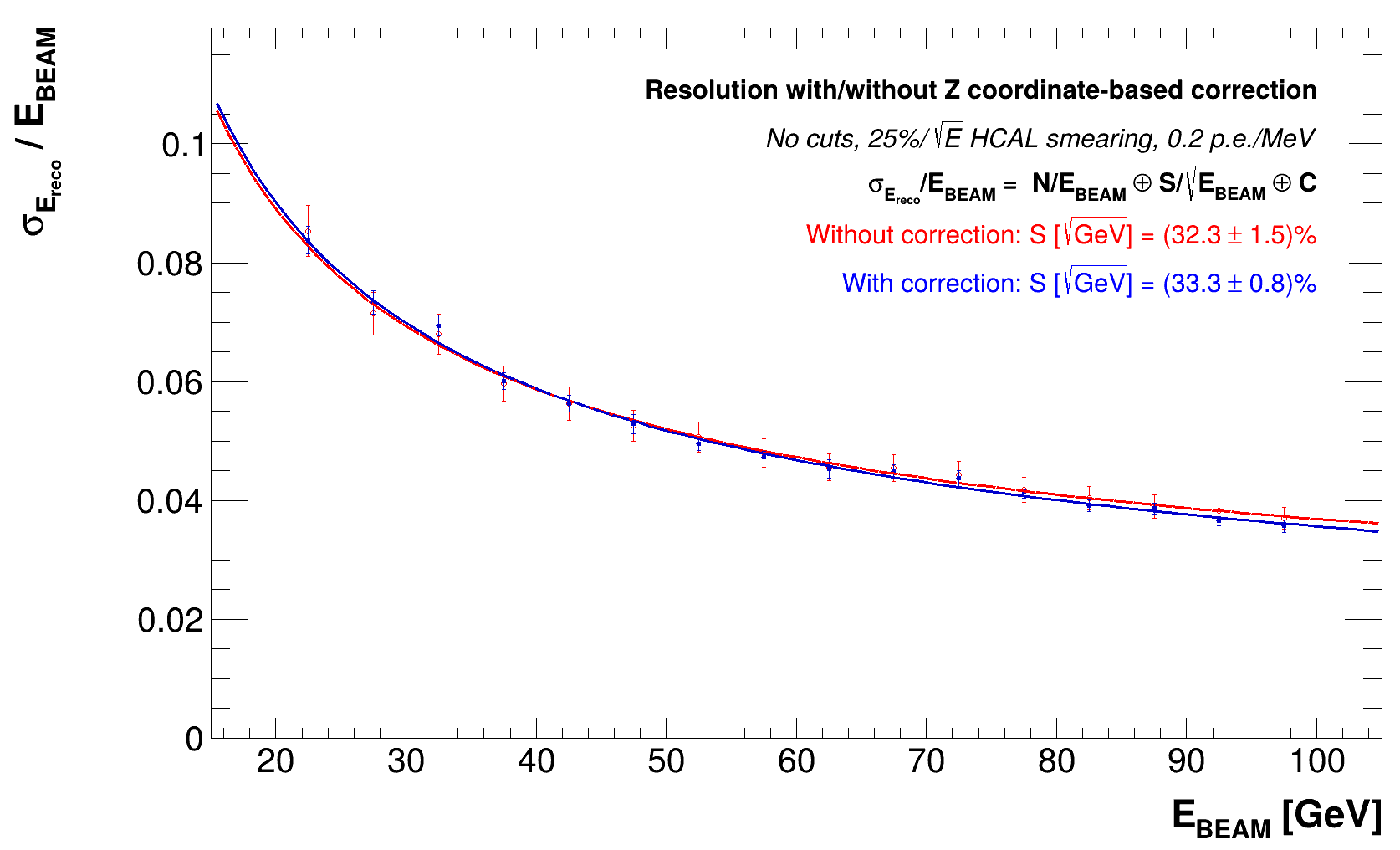}
    \caption{Energy resolutions using the RMS-based procedure, with and without the additional correction as a function of Z, applied in the same way as in Section~\ref{z}.}
    \label{fig:reso_rms_plus_gnn}
\end{figure}

Figure~\ref{fig:reso_rms_plus_gnn} compares the resolution obtained with the RMS-based correction alone and after applying the additional Z-dependent correction, suggesting that the Z-based correction is much weaker than the RMS-based one if applied sequentially and neglecting correlations.

\begin{figure}[!htbp]
    \centering
    \includegraphics[width=0.49\linewidth]{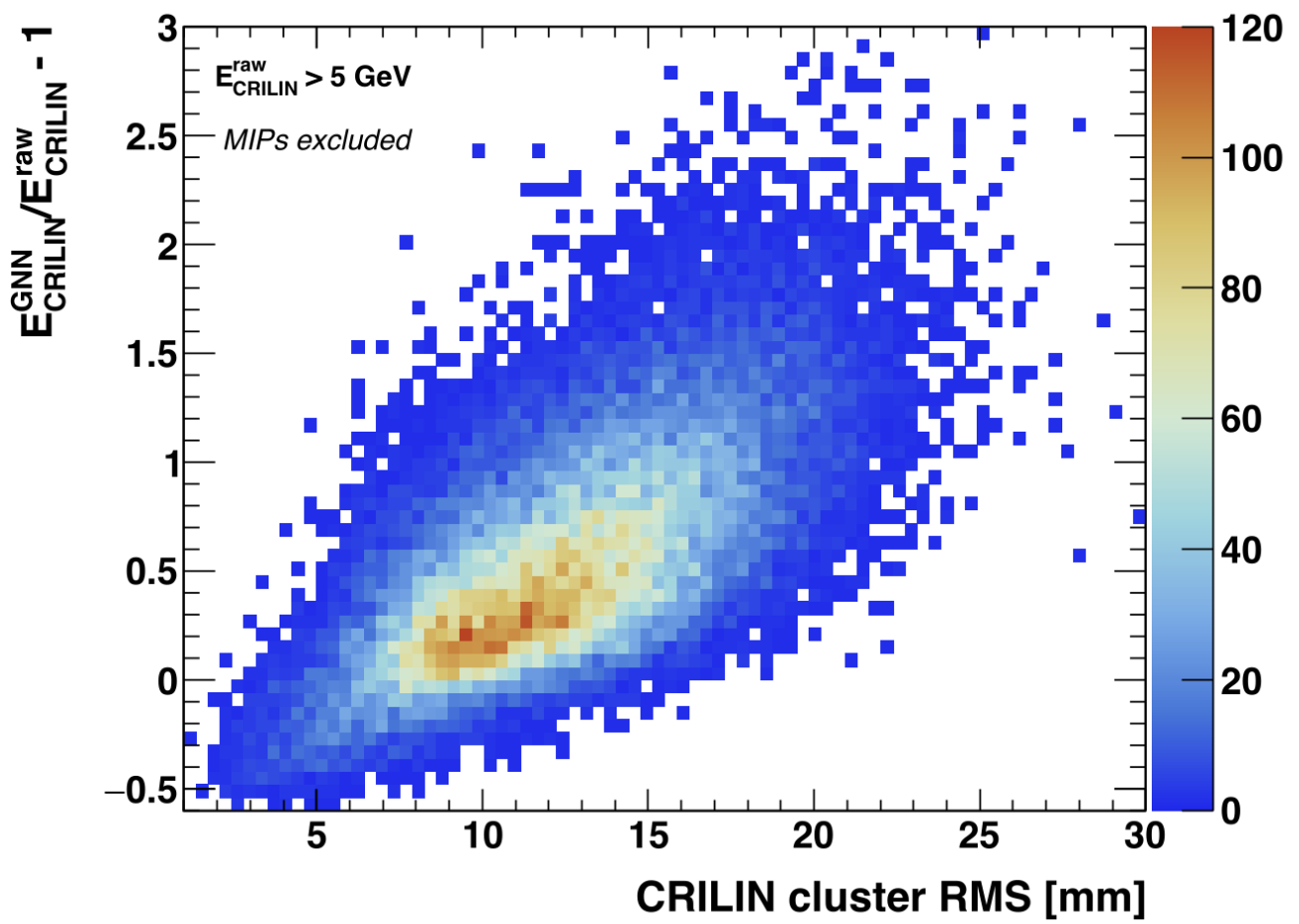}
    \includegraphics[width=0.49\linewidth]{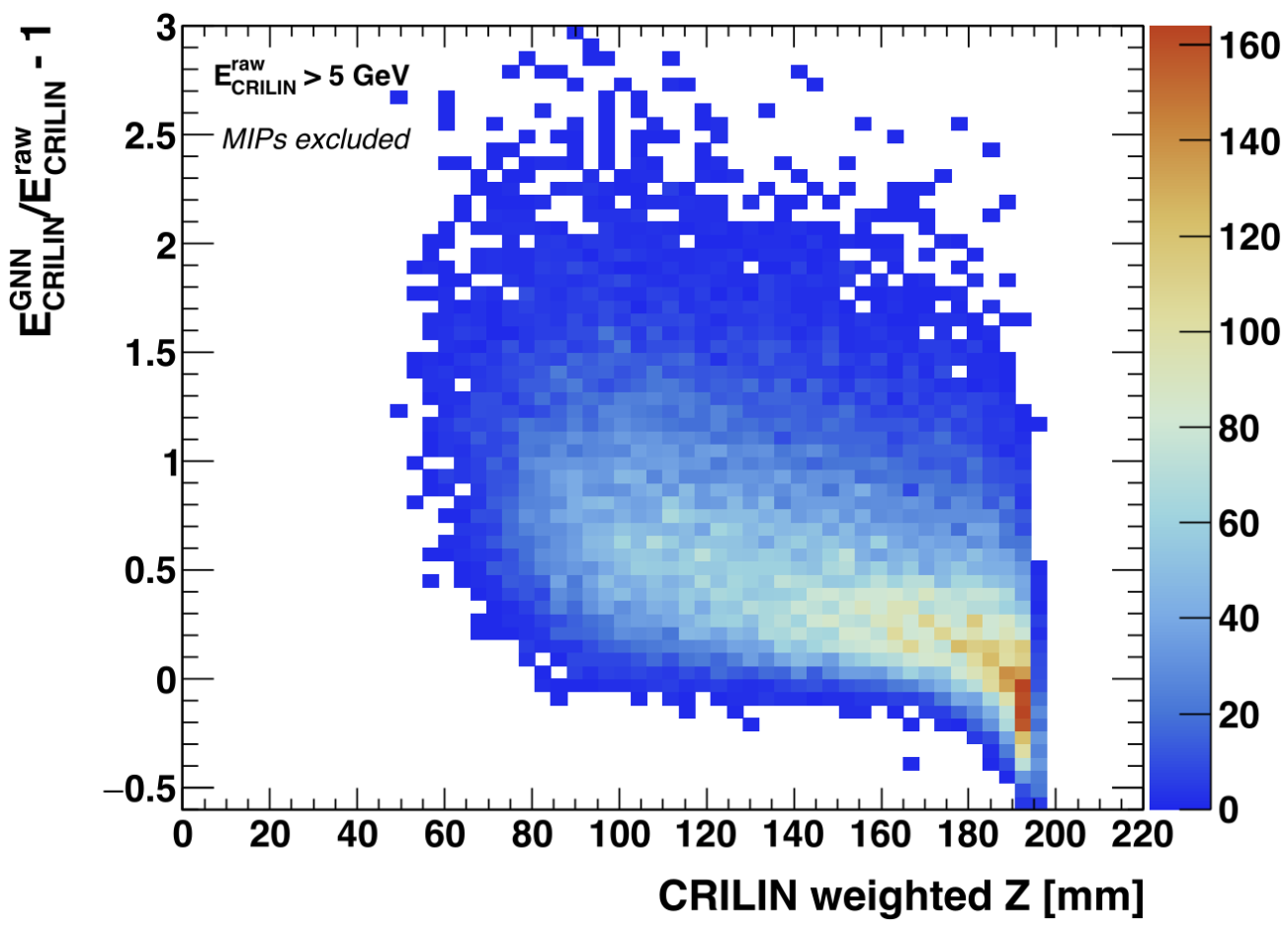}
    \caption{Joint distribution of the correction factor, defined as the ratio between the energy corrected with the GNN and the uncorrected energy, and RMS (left) or Z (right). Events corresponding to MIPs have been excluded.}
    \label{fig:gnn_z_rms}
\end{figure}

On the other hand, the usage of Z in software compensation for electromagnetic calorimeters is necessary to correctly account for edge effects at the rear end of the detector, as shown in Figure~\ref{fig:gnn_z_rms}, where the relative correction factor provided by the GNN is plotted as a function of RMS or Z.
In particular, it can be observed that, in the proximity of the rear end of the calorimeter, the GNN correction factor is close to zero, due to the smaller depth traversed by the shower.

\end{appendices}

\newpage

\bibliographystyle{elsarticle-num}
\bibliography{paper}

\end{document}